\pdfoutput=1  

\documentclass[aps,prd,superscriptaddress,floatfix,nofootinbib,preprintnumbers,
eqsecnum,twocolumn
]{revtex4-1}

\usepackage{amsmath,float,graphicx,placeins,amssymb,multirow,pgfplots,pgfplotstable}
\usepackage{tikz}
\usepackage{quantikz}
\usepackage{tikz-cd}
\usepackage{soul}
\setstcolor{red}
\setstcolor{orcidlink}

\usetikzlibrary{fit,positioning}
\usepackage{animate}

\usepackage{lipsum} 
\usepackage{comment}
\usepackage{enumerate,slashed,cancel,comment,color,bbm,graphicx,rotating,placeins,mathtools,multirow,hhline}
\usepackage[normalem]{ulem}
\usepackage{orcidlink}
\usepackage{array}
\usepackage{enumitem}
\usepackage{hyperref}
\usepackage{color}
 \hypersetup
 { 
   colorlinks,
   linkcolor={blue!80!black},
   citecolor={blue!70!black},
   urlcolor={blue!70!black}
 }

\usepackage{diagbox}
\usepackage[caption=false]{subfig}
\usepackage{ragged2e}

\usetikzlibrary{shapes.geometric,arrows}
\usetikzlibrary{calc}

\pgfplotsset{width=10cm,compat=1.9}

\graphicspath{}

\def\d{{\rm d}}
\def\ie{{\it i.e.}\/}
\def\eg{{\it e.g.}\/}

\renewcommand{\braket}[1]{{\langle #1 \rangle}}
\renewcommand{\ket}[1]{{| #1 \rangle}}

\begin{document}
\preprint{IPPP/25/81}

\title{Berry's phase on photonic quantum computers}

\def\andname{\hspace*{-0.5em}}
\author{Steven Abel\,\orcidlink{0000-0003-1213-907X}}
\email{s.a.abel@durham.ac.uk}
\affiliation{Institute for Particle Physics Phenomenology, Durham University, Durham, DH1 3LE, United Kingdom}
\affiliation{Department of Mathematical Sciences, Durham University, Durham, DH1 3LE, United Kingdom}

\author{Iwo Wasek\,\orcidlink{0009-0009-8925-5491}}
\email{iwo.a.wasek@durham.ac.uk}
\affiliation{Department of Mathematical Sciences, Durham University, Durham, DH1 3LE, United Kingdom}

\author{Simon Williams\,\orcidlink{0000-0001-8540-0780}}
\email{simon.j.williams@durham.ac.uk}
\affiliation{Institute for Particle Physics Phenomenology, Durham University, Durham, DH1 3LE, United Kingdom}

\begin{abstract}
We formulate a continuous-variable quantum computing (CVQC) algorithm to study Berry's phase on photonic quantum computers. 
We demonstrate that CVQC allows the simulation of charged particles with orbital angular momentum under the influence of an adiabatically changing $\vec{B}$ field. 
Although formulated entirely in the CVQC setting, our construction uses only passive linear-optical operations (beam splitters and phase shifts), which act identically in single-photon photonic architectures. This enables experimental realisation on the {\tt Quandella Ascella} platform, where we observe the Berry's phase phenomenon with interferometric measurement. 
We also generalise the framework to more rapid non-adiabatic evolution.  By concatenating Aharonov–Anandan cycles for opposing magnetic fields we demonstrate that one can engineer a circuit in which dynamical phases and leading non-geometric errors cancel by symmetry, leaving the intrinsically robust geometric phase contribution.
\end{abstract}
\maketitle

\def\beq{\begin{equation}}
\def\eeq{\end{equation}}

\section{Introduction} 

 Berry's phase is an important physical phenomenon that was discovered in the early days of quantum mechanics. This feature of quantum mechanics was central in several classical and quantum contexts, including Pancharatnam's early work on interference in polarised light, and in the Aharonov-Bohm effect ~\cite{Pancharatnam1956I, Pancharatnam1956II,AharonovBohm1959,HerzbergLonguetHiggins1963, LonguetHiggins1975,MeadTruhlar1979,Mead1980MolecularAB}. However its geometric nature was not fully appreciated until the developments initiated by Berry~\cite{Simon1983,Berry1984,WilczekZee1984}, 
which unified these previously disconnected observations into one simple principle that properly established its geometrical/topological origin. 

Berry's phase becomes relevant in adiabatically changing systems. Its topological character stems from the defining property of the adiabatic limit: in this limit a system changes so slowly with time that it can be described by its instantaneous energy eigenstates, which is to say the set of eigenstates $\ket{\phi_n}$ determined at time $t$ by treating the Hamiltonian as if it were constant. In such a system, provided that the energy eigenvalues remain ordered and well separated, 
\begin{equation}
    E_0(t) < E_1(t) < E_2(t) \ldots , 
\end{equation}
there is parametrically small mixing between the instantaneous energy eigenstates. Thus a system that is prepared at time $t=0$ in the $n$'th energy eigenstate, $|\psi(0)\rangle = |\phi_n(0)\rangle $, will in the adiabatic approximation evolve as 
\begin{align}
|    \psi(t)\rangle ~\approx ~ e^{i \Theta_n(t) }  e^{i \gamma_n(t) } |\phi_n(0)\rangle ~.
\end{align}

Here the phase $\Theta_n(t)$ is the straightforward non-geometric dynamical phase that one would expect to arise from the Schr\"odinger  evolution, 
\begin{equation}
    \Theta_n (t) ~=~ - \frac{1}{\hbar} \int_0^t E_n(t')\d t'~.
\end{equation}
However the phase $\gamma_n$ -- the Berry phase -- is far more interesting. It is given by 
\begin{equation}
\label{eq:gammat}
    \gamma_n (t) ~=~ i \int_{0}^t \langle \phi_n | \frac{d}{dt} |\phi_n \rangle dt~. 
\end{equation}
The geometric properties of $\gamma_n$ are manifest when it is re-expressed in terms of the parameters  that define the time-dependent Hamiltonian. If we arrange the set of $p$ parameters into a vector $ {\vec R}(t)\in {\mathbb R}^p$ which determines the instantaneous energy eigenstates $\ket{\phi_n(\vec{R})}$, then $\gamma_n(t)$ can also be written as a line integral along the curve $C(t)$ which is traversed in $\vec R$-space:
\begin{equation} \label{eq:gamma_pre}
\gamma_n(t)~=~i\int_{C(t)}
\vec {\cal A} \cdot d\vec{R}~,
\end{equation}
where $  \vec  {\cal A }$ is the Berry connection, 
\begin{equation} \label{eq:B_con}
    \vec {\cal A }~=~ i \braket{
\phi_n(\vec{R})|\nabla_{\vec{R}} |\phi_n(\vec{R})}~.
\end{equation}
From the geometric point of view, the parameter space plays the role of a base-manifold for the eigenstate bundle. Moreover, unphysical phase redefinitions of the wavefunctions, $\ket{\phi_n} \to e^{i\chi} \ket{\phi_n}$, correspond to gauge transformations of $\vec {\cal A}$, and thus the phase takes on a topological nature, becoming physically meaningful only over closed curves:
\begin{equation} \label{eq:gamma}
\gamma_n(C)~=~i\oint_C\vec {\cal A} \cdot d\vec{R}~~,
\end{equation}
for an arbitrary closed curve $C$ in $\vec R $-space.

In recent years there has been increasing interest in the Berry's phase phenomenon with the advent of quantum computing. It has been studied in several contexts \cite{ZANARDI199994,Albert2016HolonomicCV,Song2017CVGeometricPhase,Zhang2024NonadiabaticHolonomic,Ellinas2001OpticalHolonomic,Khan2021GeometricPhase,Gu2009QuantumComputingCVClusters,Tamiya2021CalculatingNonadiabaticCouplings}.
Most notably, Berry's phase based gates, and holonomic schemes more generally, promise a route towards intrinsic error resilience, offering  advantage over purely dynamic gates \cite{ZANARDI199994,Albert2016HolonomicCV}.

The purpose of this paper is to present a novel implementation of Berry's phase in the continuous-variable quantum computing (CVQC) paradigm~\cite{PhysRevLett.82.1784, Adesso2014, RevModPhys.77.513}. While previous implementations of the phenomenon in quantum computing have almost always been via loops in `control parameter space', in this work we shall perform a Hamiltonian simulation by Trotterising the real-time evolution of a physical Berry's phase system. The system we will simulate is a state with orbital angular momentum in an adiabatically evolving $\vec{B}$ field, whose Hamiltonian is the usual orbital Zeeman term $-\mu \vec{L}\cdot \vec{B}$. 

The use of the CVQC framework is especially interesting in this context because these systems operate on continuous harmonic $\hat q$ and $\hat p$ quadratures. This offers natural quantum-computing analogues of continuous Hamiltonian mechanics, enabling the straightforward implementation of continuous physical operators such as momenta and fields. CVQC platforms are therefore particularly well-suited to simulating the dynamics of a multitude of quantum systems, and they have recently gained attention for the simulation of both quantum mechanical systems and quantum field theories~\cite{Abel:2024kuv, Ale:2024uxf, PhysRevA.109.052412, Abel:2025zxb, Williams:2025vec, Abel:2025pxa}.

In the case of the Berry phase, this characteristic feature of CVQC devices allows us to study this phenomenon for orbital angular momentum, as opposed to, for example, intrinsic spin. Our main approach will be to simulate the time-dependent Hamiltonian with an evolving $\vec B$ field, using only passive Gaussian CV primitives (beam splitters and phase rotations). The Berry phase will then be extracted by interfering the resulting wavefunction against a reference arm interferometer. 

There are several reasons that this approach is attractive. Firstly, it allows us to consider the generalisation of the geometric phase to non-adiabatic systems, in the guise of the so-called Aharonov-Anandan phase, in a controlled manner. This allows us to take advantage of the robustness provided by the geometric phase's topological nature. The Berry phase, which is an exact invariant, provides a benchmark against which to test and mitigate the effect of non-adiabaticity, and errors due to other distortions such as Trotter error,  in a precise manner. Furthermore, by concatenating Aharonov–Anandan cycles for opposing magnetic fields one can engineer a geometric phase gate in which dynamical phases and leading non-geometric errors cancel by symmetry, leaving an intrinsically robust geometric phase, that is resistant to Trotter error and non-adiabaticity. 

A further advantage of this framework is that the time evolution required for the Berry's phase protocol can be implemented using single-photon states and passive optics alone, without the need for non-Gaussian operations. This enables experimental realisation of the model on current photonic hardware. An advantage of photonic devices is that they exhibit extremely low decoherence and operate without the need for extensive cryogenic cooling. In this paper, we will validate our model on the {\tt Quandela Ascella} quantum computer~\cite{Ascella}, which encodes quantum information in single photons. Although this means that {\tt Ascella} is not, strictly speaking, a continuous-variable quantum device (in the sense that one cannot make measurements of the quadrature variables themselves), it is a reconfigurable linear-optical interferometer capable of implementing arbitrary beam-splitter and phase-rotation operations on single photons, which are equivalent to passive operations on quadrature variables. Since our CVQC algorithm acts entirely within the passive linear-optical framework, these transformations are identical in both the continuous and single-photon regimes. This enables an efficient implementation of our scheme on the {\tt Quandela} platform, even though it is not a fully fledged CVQC device. In this sense for our purposes we consider {\tt Quandela Ascela} to be a platform in the CVQC paradigm, and we will refer to it as such throughout the paper. 

The paper is structured as follows. Section~\ref{sec:background} reviews Berry’s phase for orbital angular momentum in an adiabatically evolving magnetic field. Section~\ref{sec:cvqc} describes the CVQC implementation, outlining the state preparation routine, the Trotterised time-evolution circuit, and the interferometric readout method. Section~\ref{sec:AAphase} extends the analysis to the Aharonov-Anandan phase, a generalisation of the Berry phase that is applicable even when the system becomes non-adiabatic. In Section~\ref{sec:results}, we validate the model and present results from a CVQC quantum emulator and the {\tt Quandela Ascella} quantum computer. Finally, in Section~\ref{sec:conclusions} we provide a summary and conclusions.  

\section{Background: Berry's phase in orbital angular momentum}\label{sec:background}

Berry's phase is a subtle phenomenon. Indeed, in many situations it will not be manifest at all. For example, it is easy to see that in general $\gamma_n$ as it is defined in Eq.~\eqref{eq:gammat} is indeed a real phase by integration by parts: that is 
\begin{align}
     \gamma_n^* (t) ~&=~ -i \int_{0}^t \left( \frac{d}{dt}\langle \phi_n |\right)| \phi_n \rangle dt~\nonumber \\
     ~&=~ \gamma_n(t)~, 
\end{align}
by parts, since $d\langle \phi_n| \phi_n\rangle /dt = 0 $. This implies that, if an instantaneous eigenstate can be written globally as a real vector times an overall phase, then its evolution traces a contractible path in projective Hilbert space and no Berry phase can appear. In other words, Berry's phase arises only when the eigenstate develops a parameter dependent complex structure, giving a non-trivial loop in state space.
In addition a one dimensional parameter space cannot produce a Berry phase either. This can be seen from the expression in Eq.~\eqref{eq:gamma}. If the parameter space has only a single dimension, that is ${\vec R} \in {\mathbb R}$, then the closed contour  in Eq.~\eqref{eq:gamma} is trivially comprised of cancelling integrals. 

A system that {\it can} manifest a Berry's phase is a charged particle with angular momentum in a time varying  $\vec{B}$ field. As we have mentioned, often Berry's phase is considered for simple systems such as the half integer intrinsic spin contribution, but in the context of CVQC it is very interesting to examine it for particles with orbital angular momentum, defined in terms of the quadrature variables, $\vec q$ and $\vec p$: in a magnetic field \(\vec{B}\) the Hamiltonian becomes
\begin{equation}
\label{eq:ang_mom_H} H = \frac{ \vec {p}^2}{2m}  + \frac{m\omega^2 \vec q^2 }{2} -\mu \vec L\cdot \vec B ~,
\end{equation}
where $ \mu = \frac{e}{2m}$ is the gyromagnetic ratio (which we denote by $\mu$ instead of the traditional $\gamma$ to avoid confusion with the geometric phase), and where the angular momentum operators 
\begin{equation}
    \hat  L_i ~=~  \left( \vec q \times \vec p \right)_i ~,
\end{equation}
 obey the usual commutation relation 
\begin{equation}
\label{eq:Lcom}
[\hat L_i,\hat L_j] ~=~ i\hbar \epsilon_{ijk} \hat L_k~.
\end{equation}
As we shall see, a genuine CVQC device, such as a photonic quantum computer, contains all of these operations as Gaussian gates. 
Our general approach therefore will be to simulate the slow time-evolution of $\vec B(t)$ in a closed loop and measure the final phase $\gamma_n$ which is induced on a state, in addition to the dynamical phase $\int _0 ^ t E(t') \d t'$. 

\begin{figure}
\centering
\includegraphics[width=0.75\linewidth,trim=0.0cm 0.0cm 0.0cm 0.0cm,clip]{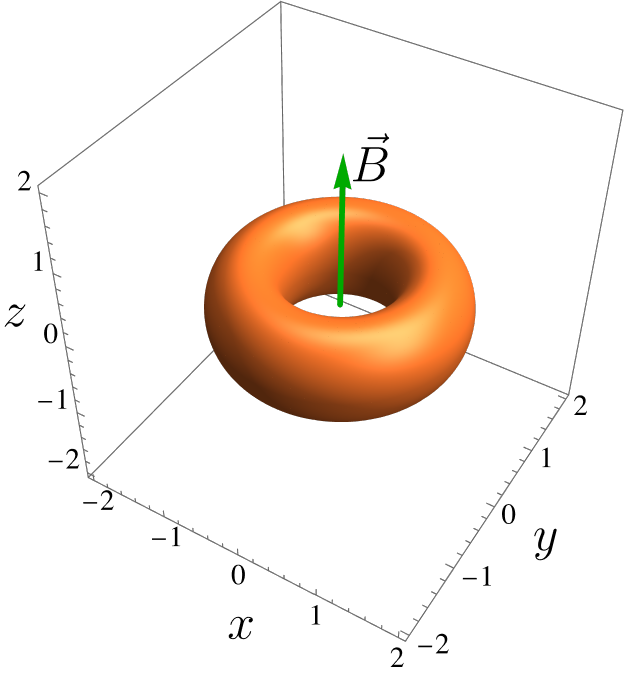}
\caption{Isosurface of the density $|\phi_{0,1,1}|^2$ (at $|\phi_{0,1,1}|^2=1/8$) of the ``donut state'' when $\vec B$ is pointing in the $\hat z$ direction.}
\label{fig:state}
\end{figure}

Let us consider how the Berry's phase is manifest in such a physical system, and postpone for the moment the question of how we are going to simulate it. For our study, it will suffice to consider systems that evolve with constant modulus for the magnetic field, such that $\vec B (t) = B\, \hat n (t)$, and for convenience let us for the moment choose a coordinate system in which the initial magnetic field is $\vec B (0) = B\, \hat z $. First, we need to choose the specific energy eigenstate that we will be subjecting to the $\vec{B}$ field.
Generally we express the states by the quantum numbers of $H_{\rm osc}$, $\vec L^2$ and $\hat L_z$, which we  denote by $n_r $, $\ell $ and $m$ respectively. By standard arguments, the energy eigenvalues can be shown to be of the form 
\begin{equation}
    E_{n_r,\ell,m}~=~ \hbar \omega ( 2 n_r + \ell + \frac{3}{2} ) - \mu B m ~.
\end{equation}
The lowest non-trivial case has $n_r=0$ and $\ell = 1$, and hence 
\begin{equation}
\label{eq:E01m}
    E_{0,1,m}~=~\frac{5 }{2}  \hbar \omega  - \mu B m ~,
\end{equation}
where the azimuthal angular momentum takes values $m = 1,0,-1$. 
Defining
\[
\alpha=\frac{m\omega}{\hbar}~,\]
 the $\ell=1$ eigenstates are given by the standard spherical harmonic decomposition to be
\begin{align}
\label{eq:l=1subspace}
\phi_{0,1,+1}(\vec q)&~\propto~ -(q_x+i q_y)\,e^{-\frac{1}{2}\alpha q^2},\quad
\nonumber \\
\phi_{0,1,0}(\vec q)&~\propto~ q_z\,e^{-\frac{1}{2}\alpha q^2},\quad
\nonumber \\
\phi_{0,1,-1}(\vec q)&~\propto~ (q_x-i q_y)\,e^{-\frac{1}{2}\alpha q^2}~,
\end{align}
where it will be convenient to express them in Cartesian coordinates. For \(\mu B>0\) the ground state within the \(\ell=1\) manifold has \(m=+1\), \ie, it is 
\begin{equation}    
\label{eq:state}
\phi_{0,1,1}(\vec q)~\propto ~(q_x+i q_y)\,e^{-\frac{1}{2}\alpha q^2}~.
\end{equation}
This lowest lying $\ell=1$ state will be the state that we will evolve adiabatically. The density for this state is shown in Fig.~\ref{fig:state}. We will often refer to it as the ``donut state''. As we shall see later, in order to single out the Berry phase it will also be useful to adiabatically evolve its friend, $\phi_{0,1,-1}(\vec q)$, or equivalently to evolve the $\phi_{0,1,1}(\vec q)$ state in the reversed $\vec{B}$ field.

\begin{figure}
    \centering
\includegraphics[width=0.8\linewidth,trim=0.0cm 0.0cm 0.0cm 0.0cm,clip]{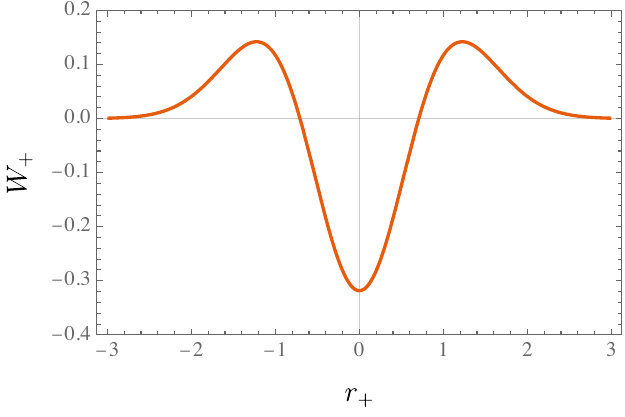}
    \caption{Wigner function $W(r_+)$ of  $\phi_{0,1,1}$, where the radius in phase space is $r_+ = \sqrt{q^2_++p^2_+}$, where the quadrature variables are $  q_\pm$ and $
    p_\mp$ are as in Eq.~\eqref{eq:quadpm}.}
\label{fig:wigner}
\end{figure}

It is instructive to present the Wigner function of these states. In principle it is six-dimensional, however four of these dimensions are somewhat trivial and it factorizes. For example, the $\phi_{0,1,1}(\vec q)$ state Wigner function is 
\begin{align}
    W_{0,1,1}~=~ W_0(q_z,p_z)\, W_0(q_-,p_-)\, W_+(q_+,p_+)  
\end{align}
where 
\begin{align}
    W_0(q,p) ~&=~ \frac{1}{\pi} e^{-(q^2+p^2)} ~,\nonumber \\
    W_+(q,p) ~&=~ \frac{1}{\pi} e^{-(q^2+p^2)} \left(2 (p^2+q^2) -1\right) ~, 
\end{align}
and where we have introduced quadrature variables in the ladder-operator basis:
\begin{align}
\label{eq:quadpm}
    q_\pm ~=~\frac{1}{\sqrt{2} } (q_x \pm p_y)~~;~~~
    p_\mp ~=~\frac{1}{\sqrt{2} } (p_x \pm q_y)~.
\end{align}
The function $W_+$ is shown in in Fig.~\ref{fig:wigner}. 

Let us now consider the adiabatic evolution of this lowest energy $\ell=1$ eigenstate. When the orientation of the magnetic field is slowly adjusted in time, the state $\phi_{0,1,1}$ changes its orientation to follow it. This induces a Berry phase which it is relatively straightforward to calculate. To do so let us write the  time-evolving state using a polarisation unit vector $\vec \varepsilon $ such that 
\begin{equation}   \label{eq:phi_eps}
\phi_{0,1,1}(\vec q)~\propto ~\vec \varepsilon \cdot \vec q \,e^{-\frac{1}{2}\alpha q^2}~,
\end{equation}
where initially we have $\vec \varepsilon = - \frac{1}{\sqrt{2}}(1,i,0)$ and $\vec B = (0,0,B)$.
If we parametrise the time-evolving magnetic field by 
\begin{equation}
    \vec B (t) ~=~ B\begin{pmatrix}
        \sin\tilde \theta(t) \cos \tilde \phi(t)\\ \sin \tilde \theta(t) \sin \tilde \phi(t)\\
        \cos\tilde \theta(t) \\
    \end{pmatrix}
\end{equation}
then the polarisation vector evolves as follows:
\begin{equation}
\label{eq:eps}
    \vec \varepsilon (t) ~=~\frac{1}{\sqrt{2} } \begin{pmatrix}
        -\cos\tilde \theta(t) \cos \tilde \phi(t) + i \sin\tilde \phi(t) \\ 
        -\cos \tilde \theta(t) \sin \tilde \phi(t)- i \cos\tilde \phi \\
        \sin\tilde \theta(t) 
    \end{pmatrix}~.
\end{equation}
One can verify that the real and imaginary components of $\vec \varepsilon $ remain orthogonal both to each other and to $\vec B$ for all time. 

\begin{figure}[ht]
  \centering

 \includegraphics[width=0.75\linewidth,trim=0.0cm 0.0cm 0.0cm 0.0cm,clip]{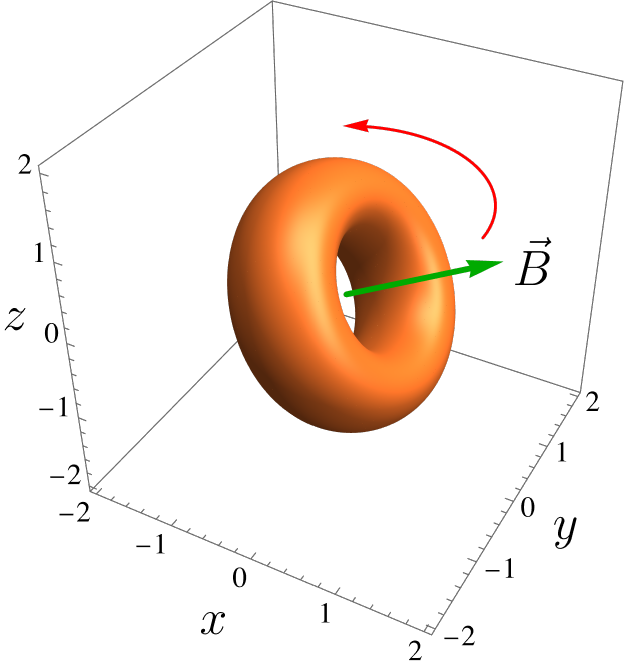}
  \caption{\label{fig:anim}Inducing Berry phase with state $\phi_{0,1,1}$, with $\vec B$ field indicated in green.}
\end{figure}

To compute the Berry phase we 
use $\tilde \phi(t) \in [0,2\pi]$ to parameterise the contour (at the risk of confusion), while the azimuthal angle is given (for consistent confusion) by $\tilde \theta(\tilde \phi )$. We stress that these angles $\tilde \phi$ and $\tilde \theta$ are parameters defining the orientation of $\vec B$ in parameter space, and are not themselves space coordinates.
Then (dropping the suffix on $\gamma$) we have 
\begin{align}
    \gamma(C) ~&=~  i \oint_C \d\vec \varepsilon \cdot  ~\langle \phi_{0,1,1} | \nabla_\varepsilon |   \phi_{0,1,1}\rangle     ~\nonumber \\
    ~&=~  i \int \d\tilde \phi ~\langle \phi_{0,1,1} | \frac{d}{d\tilde \phi}   | \tilde \phi_{0,1,1}\rangle~.
\end{align}
To evaluate the Berry connection projected along the $\tilde \phi$-direction, 
\begin{equation}
   A_{\tilde \phi} + \frac{d\tilde \theta}{d\tilde \phi} A_{\tilde \theta} ~=~ i \langle \phi_{0,1,1} | \frac{d}{d\tilde \phi}   |  \phi_{0,1,1}\rangle~,
\end{equation}
we first normalise, 
\begin{equation}   
\phi_{0,1,1}(\vec q)~= ~{\cal N} \vec \varepsilon \cdot \vec q \,e^{-\frac{1}{2}\alpha q^2}~,
\end{equation}
such that 
\begin{equation}   
\langle \phi_{0,1,1}|
 \phi_{0,1,1}\rangle ~=~ 
|{\cal N}|^2  \int \d^3 q ~ (\vec  \varepsilon ^\dagger\cdot \vec q)(\vec \varepsilon \cdot \vec q) \,e^{-\alpha  q^2}~=~1~.
\end{equation}
The integral can be performed using odd and even symmetry properties to write $(\vec  \varepsilon ^\dagger\cdot \vec q)(\vec \varepsilon \cdot \vec q) \equiv \frac{1}{3} (\vec  \varepsilon ^\dagger\cdot \vec \varepsilon ) = \frac{1}{3}$,
and then 
\begin{equation}
    |{\cal N}|^2 ~ \frac{1}{3}\int \d^3 q ~  \,e^{-\alpha  q^2}~=~1~.
\end{equation}
Again, we stress that here the angles $\theta$ and $\phi$ in the $\vec q$ integral are different from the configuration space angles $\tilde \theta $
 and $\tilde \phi$ that appear in $\vec \varepsilon$.
In a similar fashion, for the Berry connection we have 
\begin{align}   
  A_{\tilde \phi} + \frac{d\tilde \theta}{d\tilde \phi} A_{\tilde \theta}~&=~ 
i|{\cal N}|^2  \int \d^3 q ~ (\vec  \varepsilon ^\dagger\cdot \vec q)(\frac{d}{d\tilde \phi}\vec \varepsilon \cdot \vec q) \,e^{-\alpha  q^2}\nonumber \\ ~&=~i \vec  \varepsilon ^\dagger\cdot \frac{d}{d\tilde \phi}\vec \varepsilon  ~.
\end{align}
Inserting $\vec \varepsilon $ from Eq.~\eqref{eq:eps} gives 
\begin{align}
    A_{\tilde \phi} + \frac{d\tilde \theta}{d\tilde \phi} A_{\tilde \theta}~=~ \cos\tilde\theta (\tilde\phi)~,
\end{align}
which then gives 
\begin{align}
\gamma(C) ~&=~ 2 n \pi - \int \d \tilde \phi (1- \cos\tilde\theta (\tilde\phi))\nonumber \\
&=~ 2 \pi n - \Omega ~,\label{eq:gammaO}
\end{align} 
where $n\in {\mathbb Z}$ and $\Omega$ is the solid angle swept out by $\vec B$ in configuration space. 
(We are able to add $2\pi n$ to $\gamma $ as this merely corresponds to a different gauge, \eg ~$\vec\varepsilon '  = e^{i\phi} \vec\varepsilon$).
More generally for the eigenstates with $L_z=m$ we find a Berry phase of
\begin{align} 
\label{eq:berrym}
\gamma(C) ~&=~ 2 n \pi -m \Omega ~.
\end{align}
This is a rather standard kind of Berry phase result, but for the lowest non-trivial orbital angular momentum eigenstate it corresponds to the relatively involved precessing donut state, shown in Fig.~\ref{fig:anim}.

After a time $t$ the accumulated Berry phase on the lowest energy $\ell=1$  eigenstate would be 
\begin{align} 
\label{eq:gamma_gen}
\gamma(t) ~&=~ - \int_0^{\tilde\phi(t)} \d \tilde \phi (1- \cos\tilde\theta (\tilde\phi)) ~,
\end{align} 
while, because the modulus of the magnetic field is constant, the dynamical phase is simply given by  
\begin{align}
\label{eq:Theta011} 
    \Theta(t) ~&=~E_{0,1,1}t~=~  \left( \frac{5}{2} \hbar \omega -\mu B \right)~t   ~.
\end{align}

\section{Implementation in the CVQC regime}\label{sec:cvqc}

Let us now consider the implementation of such an evolution on a photonic CVQC device. As mentioned, an interesting feature of the circuit for Trotterised time evolution is that the Hamiltonian itself involves only Gaussian gates, and thus it is very straightforward to implement, as we shall see in due course. 

\subsection{Initial state preparation}\label{sec:state_prep}

As a first step we address the preparation of the initial state in Eq.~\eqref{eq:state}. 
As this state is a single-photon Fock state it is manifestly 
non-Gaussian (with the Wigner function in Fig.~\ref{fig:wigner} exhibiting the expected regions of negativity, consistent with Hudson’s theorem~\cite{Hudson1974,SotoClaverie1983}).
Thus {\it some} non-Gaussian operation is inevitable if one wishes to create the state in Eq.~\eqref{eq:state} from an initial Gaussian state. 

In a fully fledged CVQC device the creation of such a state can be achieved via a two mode squeezed vacuum state (TMSV) followed by photon counting (known as ``heralding''). To begin with let us write the state we wish to prepare as follows: 
\begin{align}
|\phi_{0,1,1}\rangle 
&~=~ \frac{1}{\sqrt2}\bigl(|1,0,0\rangle + i|0,1,0\rangle\bigr)~,
\nonumber \\
&~=~ \frac{1}{\sqrt2}(a_x^\dagger + i a_y^\dagger)|0,0,0\rangle~,
\end{align}
with qumodes ordered as \(x,y,z\), and with \(z\) being left in vacuum. This is a sum of single photon states, therefore we begin by preparing a simple single-photon state. This can be done by starting with a state consisting of an entangled ancilla mode \(A\) and intermediate mode \(C\):
\begin{align}
|\mathrm{TMSV}\rangle_{A,C}
&~=~ \sqrt{1-\lambda^2}\sum_{n=0}^\infty \lambda^n |n\rangle_A |n\rangle_C~,
\end{align}
with  $\lambda = \tanh r$, where \(r\) is the squeezing parameter, and $n$ refers to the photon number on each qumode. This is a Gaussian entangled state and is in principle easy to prepare. Performing a photon-number-resolving (PNR) detection measurement on the ancilla qumode, $A$, and selecting the  ``one photon'' outcome generates the collapse
\begin{align}
\langle 1| _A |\mathrm{TMSV}\rangle_{A,C}
&~\propto ~|1\rangle_C~.
\end{align}
Thus, mode $C$ is approximately collapsed to a single photon state. 
The probability of successfully detecting one photon in the ancilla is
\begin{align}
P_{\mathrm{herald}} &~=~ (1-\lambda^2)\lambda^2~,
\end{align}
so that finite squeezing (\(\lambda\) not infinitesimal) is crucial, and implies higher-photon contamination in the unheralded runs. However, by conditioning on the single-photon detection and choosing moderate \(\lambda\) one obtains high fidelity. By contrast, in single-photon devices such as {\tt Quandela Ascella} the preparation is much easier, as the individual photons are injected into the linear optical network from quantum-dot sources, and arbitrary configurations of single-photon states can be delivered to the photonic chip as desired.  

This completes the entire non-Gaussian content of our discussion. Next, let us introduce a vacuum mode $V$  (which will become the $y$ mode) and interfere modes \(C\) and \(V\) with a balanced beam splitter as follows:
\begin{align}
a_x^\dagger &= \frac{a_C^\dagger + a_V^\dagger}{\sqrt2}~, 
~~
a_y^\dagger = \frac{a_C^\dagger - a_V^\dagger}{\sqrt2},
\nonumber \\
|1\rangle_C|0\rangle_V
&~\longrightarrow \frac{1}{\sqrt2}\bigl(|1,0\rangle + |0,1\rangle\bigr)_{x,y}.
\end{align}
Finally a \(\pi/2\) phase-shift is applied to the $y$ mode, \ie $ R_y(\pi/2)$, which inserts the required factor of $i$:
\begin{align}
\frac{1} {\sqrt 2} \bigl(  |1,0\rangle + |0,1\rangle\bigr) _{x,y}
&\xrightarrow{R _y(\pi/2)}
\frac{1}{\sqrt2}  \bigl(|1,0\rangle +   i |0,1\rangle\bigr) _{x,y}
~.
\end{align}
Including the inert $z$-vacuum factor, this is our desired starting state. The  procedure above would be applicable for state preparation on a genuine photonic device and is straightforward to implement. 

Thus once the ``heralding'' step has been carrier out and a single photon state prepared, one need only implement
\begin{align}
\label{eq:stateprep}
|\phi_{0,1,1}\rangle 
&~=~ R_y(\pi/2)\,\mathrm{BS}_{x,y}\,|1,0,0\rangle~
\end{align}
to prepare our state: \ie, we start with a single photon in mode \(x\), and then apply the same balanced beam splitter between \(x\) and \(y\), and phase shift on mode \(y\) by \(\pi/2\), as described above.

\subsection{Gadget for real-time evolution}

Having prepared the initial state, the rest of the discussion proceeds in a Gaussian way, and will ultimately yield a passive circuit that is common to both single photon and genuine CVQC machines. In order to evolve the state in real time we will use Trotter-Suzuki evolution. Therefore, we next need to configure a Trotter ``gadget'' that will approximate the evolution of the system through a small time $dt$. 
We will consider unit mass and angular frequency, $m=\omega = \hbar = 1$, thus our Hamiltonian takes the explicit form  
\begin{align}
\label{eq:H_final}
H &~=~ \frac{1}{2}\sum_{i=x,y,z}(p_i^2+q_i^2)-\mu\,\vec{B}\cdot\vec{L}~,
\end{align}
with components \(\hat L_x=q_y p_z-q_z p_y\) {\it etc}. The exact evolution operator over a time \(T\) is \(U(T)=e^{-i H T}\), and the Trotter approximation to it can be implemented using the ``gadget'' 
\begin{align}
G^{(1)}(dt)
&~=~ 
e^{-i \frac{dt}{2} H_{\mathrm{osc}}}
\,e^{i\mu\,\vec{B}\cdot\vec{L}\,dt}
\,e^{-i \frac{dt}{2} H_{\mathrm{osc}}}~,
\end{align}
where 
\begin{align}
U(T) ~\approx ~ G^{(1)}(dt) ^{T/dt} ~.
\end{align}
As we have already mentioned all terms in the operator $G^{(1)}$ can be implemented with Gaussian gates, for example using 
\begin{align}
R_i(\phi) &~=~ e^{i\phi\,\hat a_i^\dagger \hat a_i}
&&\text{(rotation on mode \(i\))},\nonumber \\
\mathrm{C_X}_{j\to k}(s) &~=~ e^{-i s\,q_j p_k}
&&\text{(controlled-}X\text{ from \(j\) to \(k\))}\nonumber ~.
\end{align}
It is straightforward to verify that (defining $\varepsilon _{ijk}$ to be the usual Levi-Cevita symbol)
the evolution operation becomes
\begin{widetext}
\begin{align}
G^{(1)}(dt)
~\approx ~
\left[\prod_{i=x,y,z} R_i\!\left(-\frac{dt}{2}\right)\right]
&
\, 
\left[
\prod_{i,j,k} \mathrm{C_X}_{i\to j}(-\varepsilon_{ijk} \mu B_k dt)\,
\right] 
\,\left[\prod_{i} R_i\!\left(-\frac{dt}{2}\right)\right]~.
\end{align}
The error is ${\cal O}((H dt)^2)$ per step, and the error accumulated after $N = T/dt$ steps is order $H^2 T dt$. The operator $G^{(1)}(dt)$ corresponds to the circuit diagram in Fig.~\ref{fig:gates}. Note that in these gates the parameters $B_i$ are themselves functions of $t$. Our entire time evolution will consist of a set of $T/dt$  such ``gadgets'' with their $t$ dependent parameters, where $T$ is the total time.   

A second order (Strang) approximation can be achieved by doubling the configuration and halving the time. That is $U(T)\approx G^{(2)}(dt)^{T/dt}$ where 
\begin{align}
    G^{(2)}(dt) ~=~ G^{(1)}(\frac{dt}{2})\cdot (G^{(1)}(\frac{dt}{2})) ~~.
\end{align}
This incurs a smaller error of ${\cal O}((H dt)^3)$ per step, accumulating an error of order $H^3 T dt^2$ after $N = T/dt$ steps. The process can be iterated, with each improvement in the approximation costing twice as many gates, and suppressing the error by an extra factor $dt$. For this study we will use the Strang approximation.

The configuration above is Gaussian and may be implemented in \eg\/ {\tt Strawberryfields}~\cite{Killoran2019strawberryfields} with no need for non-linear gates. 
In practice though, despite being Gaussian, controlled-X gates are hard to implement because they are not passive (\ie, they do not conserve photon number). However, upon inspection the total Hamiltonian in Eq.~\eqref{eq:ang_mom_H} {\it is} passive, and indeed an alternative method for implementing the evolution is available, in which the components of $e^{i \mu B_i L_i dt}$ are generated using beam splitters. In terms of creation and annihilation operators a beam splitter on modes $i,j$ generates
\begin{align}
    \mathrm{BS}_{i,j}(\theta ,\phi)  ~=~ \exp[\theta (e^{i\phi} \hat a_i \hat a _j^\dagger - e^{i\phi} \hat a_j \hat a _i^\dagger) ]~.
\end{align}
Thus, since $\hat q_i = \frac{1}{\sqrt{2}} (\hat a_i + \hat a^\dagger _i ) $ and $i \hat p_i = \frac{1}{\sqrt{2}} (\hat a_i - \hat a^\dagger _i ) $,
we straightforwardly find 
\begin{align}
    e^{i\mu \vec B \cdot \vec L  dt }~=~ \mathrm{BS}_{y,z} (dt \, \mu B_x,0)~\mathrm{BS}_{z,x} (dt \,\mu B_y,0)~\mathrm{BS}_{x,y} (dt \,\mu B_z,0)~,
\end{align}
and hence an alternative and entirely passive Trotter-Suzuki gadget is the following:
\begin{align}
G^{(1)}(dt)
~\approx ~
\left[\prod_{i} R_i\!\left(-\frac{dt}{2}\right)\right]
&
\, 
\left[
\prod_{x,y,z}^{\circlearrowright}
\mathrm{BS}_{x,y}(\mu B_z dt,0)\,
\right] 
\,\left[\prod_{i} R_i\!\left(-\frac{dt}{2}\right)\right]~.
\end{align}
The corresponding circuit is shown in Fig.~\ref{fig:gates2}.
Of course, in simulations the distinction is not important and either gadget works  (although the passive gadget is significantly faster to simulate and can also be implemented on {\tt Quandela}). At the time of writing, a passive gadget is necessary for the practical implementation on {\tt Quandela}. We then embed this gadget into the full circuit toolchain, shown in Fig.~\ref{fig:berry_interferometer}, to carry out the Trotterised time evolution.

\tikzset{
  gate/.style={draw, rectangle, minimum width=1.cm, minimum height=0.7cm, font=\footnotesize, inner sep=3pt, outer sep=0pt},
  ctrl/.style={circle, fill=black, inner sep=1.5pt},
  wire/.style={thick},
  label/.style={font=\footnotesize}
}

\begin{figure*}
\includegraphics[width=\textwidth]{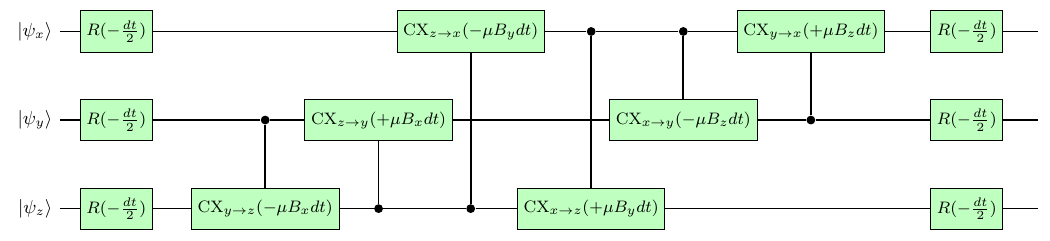}
\caption{\label{fig:gates} The Trotter-Suzuki  ``gadget'' $G^{(1)}(dt)$ for a single time step through time $dt$, to generate real time evolution under the influence of the Hamiltonian in Eq.~\eqref{eq:H_final}. The second order Strang ``gadget'' is given by halving the time interval and doubling the gates,  $S(dt) = G(\frac{dt}{2}) \cdot (G(\frac{dt}{2}))^T$.}
\end{figure*}

\begin{figure*}
\includegraphics[scale=1]{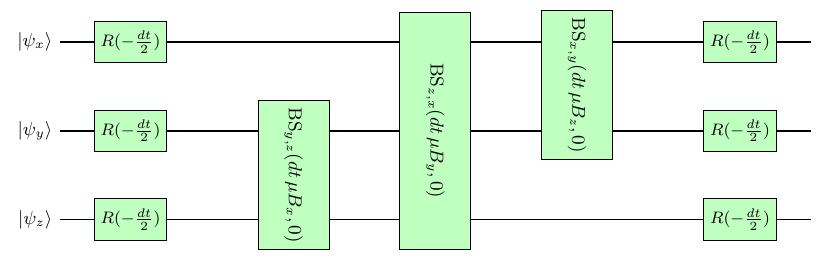}
\caption{\label{fig:gates2} The Gaussian Trotter-Suzuki  ``gadget'' $G^{(1)}(dt)$ for a single time step through time $dt$, to generate real time evolution under the influence of the Hamiltonian in Eq.~\eqref{eq:H_final}. Here the pairs of CX gates in Fig.~\ref{fig:gates} that generated $e^{i \delta t \mu B_i L_i }$ are replaced by beam splitters.}
\end{figure*}

\begin{figure*}[ht]
\centering
\includegraphics[width=\textwidth]{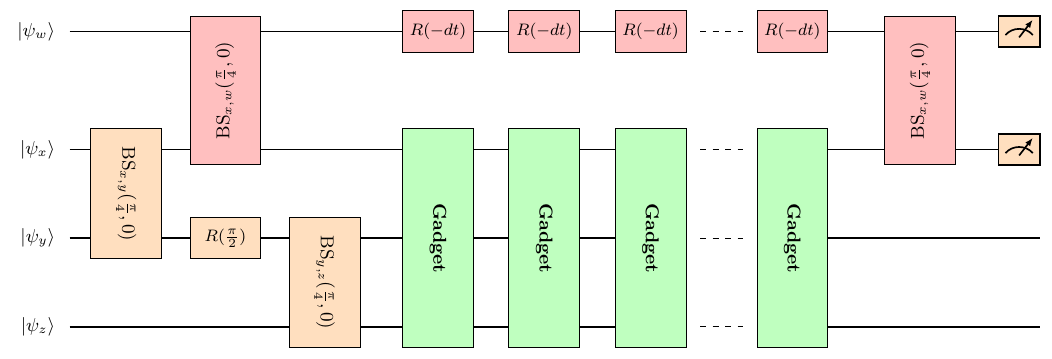}
\caption{Four‐mode interferometer: an initial ``donut state'' is prepared  (left) by the orange gates, in superposition with an inert $m=0$ eigenstate, and then  split to the reference qumode, $w$, using a balanced beam splitter. A series of Trotter-Strang step gadgets (\ie\/ the configuration in Fig.~\ref{fig:gates}) evolve the donut state on modes $x,y,z$ through any desired path in $\vec B$ parameter space, while the reference state on mode $w$ is evolved with rotation gates such that it acquires the same dynamical phase on the $w$-qumode. Finally modes $x$ and $w$ are recombined for interferometric read‐out. }
\label{fig:berry_interferometer}
\end{figure*}

\end{widetext}

It is worth remarking on the resource scaling in the particular problem that we are studying here. We begin with a fixed non-Gaussian input state
whose Wigner function \(W_{\mathrm{in}}(\vec q ,\vec p)\) is known. However, the subsequent gates in the circuit which implements the time evolution will be either controlled-X gates or beam splitters or rotations, all of  which are Gaussian unitaries. Such unitaries act as linear symplectic transformations on phase space, so if \(S\in\mathbb{R}^{m\times m}\) is the total symplectic matrix for the composed Gaussian evolution (with \(m=2\times\) number of modes $=6$), then the final Wigner function is of the form
\begin{align}
W_{\mathrm{out}}(\vec q,\vec p)
&~=~ W_{\mathrm{in}}(S^{-1}(\vec q,\vec p))~.
\end{align}
Thus, in the SHO type Hamiltonian of Eq.~\eqref{eq:H_final} that we are considering, no new non-Gaussianity is generated, and the initial non-Gaussian feature is carried along by a mapping of coordinates. This is in fact another way to see that the net measurable effect of such an evolution will amount to a simple Berry phase. 

This is why, ultimately, on the real photonic device, the entire post-injection circuit can consist of passive linear optics (beam splitters and phase shifts that conserve total photon number), and therefore the single-photon amplitude can be propagated very cheaply (less so for an an-harmonic oscillator Hamiltonians).
Likewise, in order to perform a simulation one can in for example {\tt StrawberryFields} work in a highly truncated Fock basis, because in principle the Gaussian gates should not spread amplitude to higher photon number states. This picture is of course aligned with the goal of adiabaticity. Thanks to the adiabatic theorem we know that (assuming we can perform our analysis correctly) any spread to higher Fock states should remain exponentially suppressed, and we can anticipate that the analysis for this kind of problem will continue to be resource efficient. 

\begin{figure}
    \centering
\includegraphics[width=0.55\linewidth,trim=0.0cm 1.3cm 0.cm 0.cm,clip]{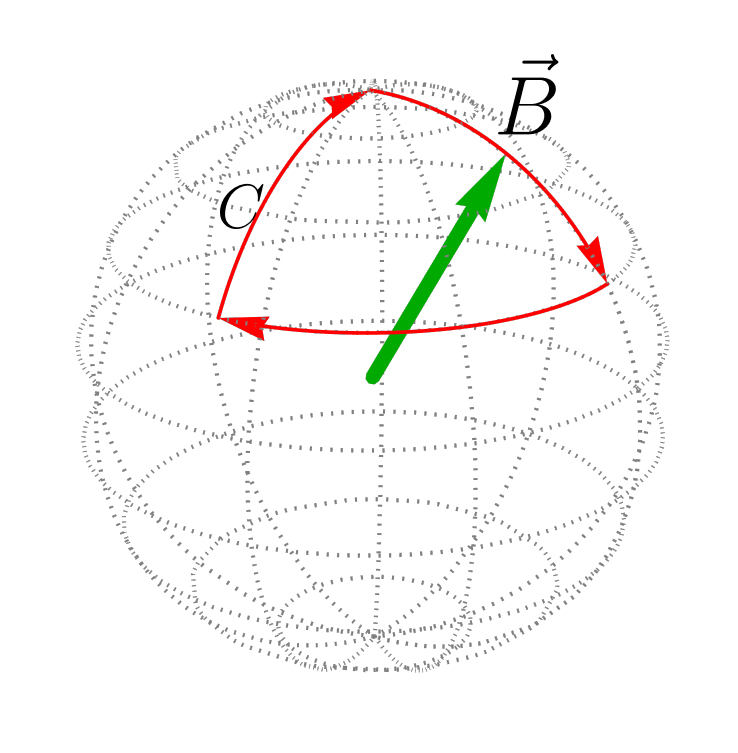}
\caption{The closed contours $C$ used for this study (in red). The $\vec B$ field starts pointing along the $\hat z$-axis and is tilted down to an azimuthal angle $\theta_0$, rotated by $\phi_0$ and tilted back up to its starting position. The solid angle swept out by $\vec B$ is $\Omega = \phi_0 (1-\cos\theta_0)$.}
\label{fig:contour}
\end{figure}

All that remains for the evolution then is to choose a time-dependent closed curve in parameter space,  $\vec B (t)$, which will generate a Berry phase. 
For this purpose, it is convenient to consider the path shown in Fig.~\ref{fig:contour}, which is composed of three segments each taking time $T_i$ (with the total time being $T=T_1+T_2+T_3$). 
The path begins with the $\vec B$ field point along the $ z$-axis and then tilts down to azimuthal angle $\theta _0$ during time $T_1$, sweeps through angle $\phi_0 $ during time $T_2$, and finally returns to point along the $z$-axis during time $T_3$. The modulus $B=|\vec B|$ will be kept constant at all times, so that the energy eigenvalues also remain constant. 
Thus the path can be  
defined by the following time-dependent values for the unit direction vector $\hat n(t)$ of $\vec B (t) = B \hat n (t)$:
\begin{equation} 
\hat n(t) = \begin{cases}
\bigl(\sin\tilde \theta_1 ,\,0,\,\cos\tilde \theta_1\bigr);
& \hspace{-0.1cm}  t \in [0, T_1]\\[8pt]
\bigl(\sin\theta_0\cos\tilde \phi_2,\sin\theta_0\sin\tilde \phi_2,\cos\theta_0\bigr);
& \hspace{-0.1cm} t\in [T_1, T-T_3]\\[8pt]
\bigl(\sin\tilde\theta_3 \cos \phi_0,\sin\tilde\theta_3 \sin\phi_0 , \cos\tilde\theta_3 \bigr);
& \hspace{-0.1cm} t\in [T - T_3 , T]~
\end{cases}
\label{eq:angles_T}
\end{equation}
where
\begin{align} 
\tilde \theta_1(t)&~=~ \frac{t}{T_1} \theta_0\nonumber \\
\tilde \phi_2(t)&~=~ \frac{t-T_1}{T_2}\phi_0 \nonumber \\
\tilde \theta_3(t)&~=~ \frac{T-t}{T_3}\theta_0 \end{align}
with 
\begin{align}
\theta_0 = \arccos\, \Bigl(1 - \frac{\Omega}{\phi_0}\Bigr)~,
\end{align}
such that \( \Omega = \phi_0\,(1 - \cos\theta_0 )\).
Thus by Eq.~\eqref{eq:gamma_gen} the accumulated Berry phase is 
\begin{align}
\label{eq:gamma_C}
\gamma(C) ~=~ -\, \phi_0\,(1 - \cos\theta_0)~.
\end{align}

We also note that as Berry's phase depends on the solid angle subtended by the contour, if we were to consider a different contour that encloses the same area (and thus the same solid angle) on the surface of the sphere, it must give the same Berry's phase. Hence, we will also consider the anti-polar path, defined by the time-dependent values for the unit direction vector $\hat{m}(t)$ of $\vec{B}(t)=B\,\hat{m}(t)$, where we have $\hat{m}(t)=-\hat{n}(t)$, with $\hat{n}(t) $ defined in \eqref{eq:angles_T}. This clearly has the same solid angle as the path above, but the $\phi_{0,1,-1}$ state is now the lowest energy eigenstate, rather than the $\phi_{0,1,1}$ state. In other words, taking $\phi_{0,1,1}$ around the anti-polar path is equivalent to taking  $\phi_{0,1,-1}$
around the original polar path. Given Eq.~\eqref{eq:berrym} it is clear that one can add the two in order to expose the Berry phase. 
This will prove to be very useful later in the paper.

\subsection{Circuit for interferometric measurement of Berry's phase }

There are several approaches one could take for observing the Berry phase which is induced by the Hamiltonian evolution of the previous section. The most comprehensive procedure would be to reconstruct the entire wavefunction of the (almost pure) end state with wavefunction tomography. That is, after the evolution we would measure the Fock probabilities, and also insert beam-splitters to interfere the one photon modes and measure their  relative phases. However, this would require at least five repeat runs, and is somewhat excessive since we only wish to measure the one phase. A much more efficient approach, which we will use here, is to instead split the incoming state into a ``signal'' state and a ``reference'' state, and then interfere them once the signal state has been traversed around its closed path in parameter space. This is in essence equivalent to performing an Aharonov-Bohm like experiment within our simulation. 

The expanded configuration for such a procedure is shown in Fig.~\ref{fig:berry_interferometer}. As well as the Berry circuit on modes $x,y,z$, we add a fourth qumode $w$, which will carry the ``reference'' state. First, on the $x,y,z$-qumodes we prepare the continuous‐variable “donut state'' by creating a single excitation in the $x–y$ subspace and imprinting the relative phase as per Eq.~\eqref{eq:stateprep} and as 
 indicated by the orange gates in Fig.~\eqref{fig:berry_interferometer}.
This leaves us with 
\begin{align}
\lvert\psi_{\rm donut}\rangle
&~=~\frac{1}{\sqrt{2}}\Bigl(\lvert1,0,0\rangle + i\,\lvert0,1,0\rangle\Bigr)~,
\label{eq:donut}
\end{align}
on the $x,y,z$-qumodes. Next, we split the $x$-qumode into a signal rail and a reference rail (the  $w$-qumode) via a balanced 50-50 beam splitter. This copies the $x$ component of \(\lvert\psi_{\rm donut} \rangle\) onto the reference arm so that the total four-qumode state becomes 
\begin{align}
\label{eq:bs1}
\ket{\psi}
&~=~
\frac{1}{2}
\Bigl(
\ket{1,0,0,0} + 
\ket{0,0,0,1} \Bigr) + \frac{i}{\sqrt{2}} \ket{0,1,0,0} ~.
\end{align}
It is easy to see that adopting the state at this juncture as our initial state would result in a complicated interference from which it  would be very difficult to extract the Berry phase. Indeed 
the additional $|0,1,0\rangle $ piece (dropping the inert $\ket{0}_w$ vacuum factor) may be decomposed using  Eq.~\eqref{eq:l=1subspace} into the sum of $m=\pm 1 $ eigenstates:
\begin{align}
    \ket{0,1,0}_{xyz}~=~ \frac{i}{\sqrt{2} } (\ket {\phi_{0,1,1}}+ \ket{\phi_{0,1,-1} })~,
\end{align}
with the phase chosen as in Eq.~\eqref{eq:l=1subspace}.
The dynamical phases are different for these two components because from Eq.~\eqref{eq:Theta011} they have energies 
\begin{align}
\label{eq:Theta011_second} 
 E_{0,1,m}~&=~  \frac{5}{2} \hbar \omega -m\mu B  ~.
\end{align}
Adding the vacuum energy of the reference $w$-mode, and setting $\hbar=\omega =1$, the dynamical phases are thus $\Theta_\pm = E_\pm T =  3T\mp \mu B T$ for the $m=\pm 1$ components of the wavefunction when the photon is on the $x,y,x$-qumodes, and simply $\theta_0=3T$ when the photon is either on the $w$-qumode or in the $m=0$ eigenstate on the $z$-qumode. Meanwhile 
the Berry evolution acts on the single‐photon subspace with phases $\gamma=-m \Omega$ as in Eq.~\eqref{eq:berrym}. Thus, if we were to work with the state in Eq.~\eqref{eq:bs1}, there would be contributions to the probabilities with both different dynamical phases and different Berry phases. 

To simplify the interferometric analysis therefore, we next add a third balanced beam splitter into our state preparation which acts on the $y$ and the $z$ qumodes. This sends
\begin{align}
    \ket{0,1,0,0} ~\longrightarrow ~ \frac{1}{\sqrt{2}} \left(
    \ket{0,1,0,0}+\ket{0,0,1,0}\right) ~,
\end{align}
removing some of the $|0,1,0,0\rangle$ amplitude and putting it onto the $z$ qumode. Thus at the end of this initialization we have the following state: 
\begin{align}
\label{eq:init}
\ket{\psi(0)}
&~=~
\frac{1}{2}
\Bigl(
\ket{1,0,0,0} + 
i \ket{0,1,0,0} \Bigr) 
\nonumber \\
& ~~~~~~~~~+ \frac{1}{2} \Bigl( 
 \ket{0,0,0,1}+ i \ket{0,0,1,0} \Bigr)~.
\end{align}
At this point we may perform Trotterised evolution on the signal arm by repeated application of the gadget of Fig.~\ref{fig:gates} on the $x,y,z$-qumodes (with its appropriate time-evolving $\vec B(t)$ value),
while applying identical free‐oscillator rotations \(R(-{d t})\) 
on the reference qumode (indicated in red in Fig.~\ref{fig:berry_interferometer}).

 It is rather simple to work out the expected evolution of the state. The first line of 
the initial state in Eq.~\eqref{eq:init} is our $m=+1$ mode. 
Thus, after having completed a closed circuit $C$ in time $T$, this component of the state returns to its starting position with a dynamical phase $\Theta = 3T-\mu B T$ and with a Berry phase of $\gamma = -\Omega$. The remaining pieces on the second line of Eq.~\eqref{eq:init} both evolve without a Berry phase and with a dynamical phase of $\Theta = 3T$. 
Thus, the final state at time $T$ is expected to be 
\begin{align}
\label{eq:final}
e^{3iT}\ket{\psi(T)}
&~=~
\frac{e^{i\Delta} }{2}
\Bigl(
\ket{1,0,0,0} + 
i \ket{0,1,0,0} \Bigr) 
\nonumber \\
& ~~~~~~~~~+ \frac{1}{2} \Bigl( 
 \ket{0,0,0,1}+ i \ket{0,0,1,0} \Bigr)~,
\end{align}
where we have defined the single combined angle 
\begin{align} 
\Delta(T) ~=~ \mu B T + \gamma(C) ~,
\end{align}
and where the accumulated Berry phase $\gamma = -\Omega$ along the path defined by Eq.~\eqref{eq:angles_T} is given by Eq.~\eqref{eq:gamma_C}. 

The final ingredient in the circuit of Fig.~\ref{fig:berry_interferometer} is of course the measurement itself. 
To do this, 
following the time evolution the $x$ and $w$-qumodes are recombined with a fourth balanced beam splitter (shown in red in Fig.~\ref{fig:berry_interferometer}), which yields 
\begin{align}
e^{3i T} \ket{\psi}_{\rm out}
&~~=~~
\frac{1}{\sqrt{2}}\left(  \frac{e^{i\Delta}-1}{2}  \right) \ket{1,0,0,0} \nonumber \\
&~~~+~~\frac{1}{\sqrt{2}}\left(  \frac{e^{i\Delta}+1}{2}  \right) \ket{0,1,0,0} \nonumber \\
&~~~+ ~\frac{1}{\sqrt{2}}\,\ket{\mbox{\,photon on mode $y$ or $z$}}~.
\end{align}
We conclude that the relative probabilities of measuring a photon on the ``signal'' $x$-qumode or the ``reference'' $w$-qumode are given by 
\begin{align}
    P_{\rm sig}~=~ \frac{1}{2}\left( 1+\cos \Delta \right);~    
    P_{\rm ref}~=~ \frac{1}{2}\left( 1-\cos \Delta \right)~.
\end{align}
Thus, our careful state preparation has yielded the ideal Mach–Zehnder interference fringe.

The Berry phase may now be inferred by determining $\Delta$ at time $T$ by photon counting on the $x$ and $w$-qumodes. 
In detail, a photon number count of $N_{\rm sig}$ on the $x$-qumode and  $N_{\rm ref}$ on the $w$ qumode corresponds to 
\begin{align}
    \gamma(C) ~\approx~ - \cos^{-1} \left( \frac{N_{\rm sig} - N_{\rm ref}}{N_{\rm sig} + N_{\rm ref}}\right)  - {\rm arg}(e^{i B\mu T}) ~,
    \label{eq:master} 
    \end{align}
where we consider the case where $N=N_{\rm ref}+N_{\rm sig}$ to be very large (\ie, we run the experiment $N$ times). 

Note that in this expression there is a sign ambiguity in the arc-cosine in addition to the ambiguity under addition of $2\pi$. Although this can be inferred and compensated by hand, it is worth mentioning that there is a more sophisticated version of the experiment in which the ambiguity is removed by performing two runs, with the second run including a phase of $\pi/2$ on the reference arm. This produces $\sin\Delta $ instead of $\cos\Delta $ and if for example $T=2\pi n /\mu B$, then one can determine the Berry phase as 
\begin{align}
    \gamma(C) ~\approx~ - \tan^{-1} \left( \frac{N'_{\rm sig} - N'_{\rm ref}}{N'_{\rm sig} + N'_{\rm ref}}~\frac{N_{\rm sig} + N_{\rm ref}}{N_{\rm sig} - N_{\rm ref}}\right)  ~,
    \label{eq:master2} 
    \end{align}
where the primed quantities are the measurements made for the second run. 

In practice we use the simpler version and take Eq.~\eqref{eq:master} as our master equation to accompany the interferometer circuit in Fig.~\ref{fig:berry_interferometer}. For simplicity one could for example take $T=2\pi n /\mu B$ to expose $\gamma$. 
For such values {\it any} photon observed on the reference $w$-qumode would indicate a non-zero Berry phase. 
However, it is also interesting to vary $T$ continuously to check the correct time-dependence.

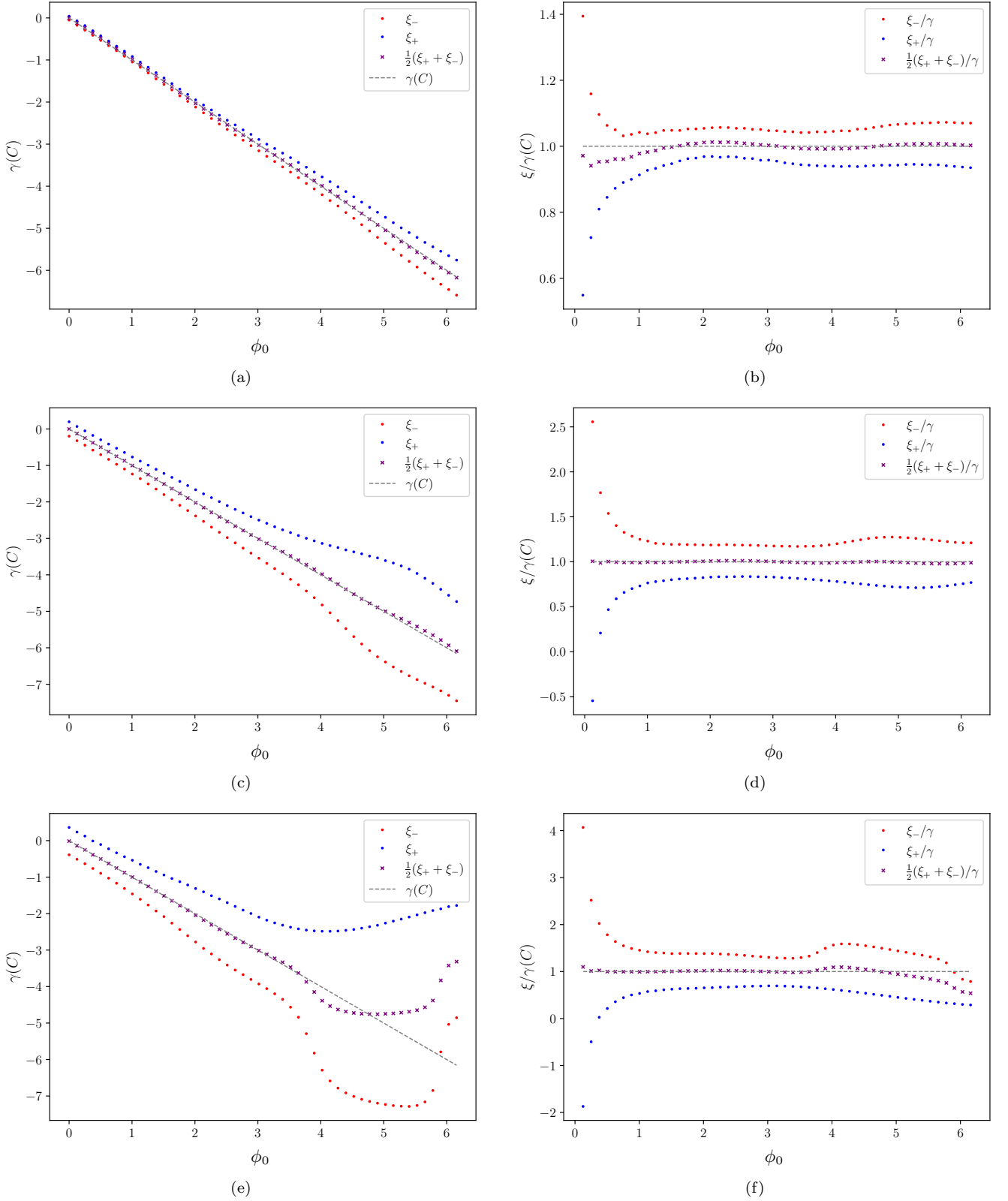
\begin{figure*}[t]
    \centering

    \subfloat[\label{fig:Berry's phase graph phi a}]{
        \begin{tikzpicture}
          \node[anchor=south west, inner sep=0] (img)
            {\includegraphics[width=0.48\linewidth,
                trim=0 0.8cm 0 0, clip]{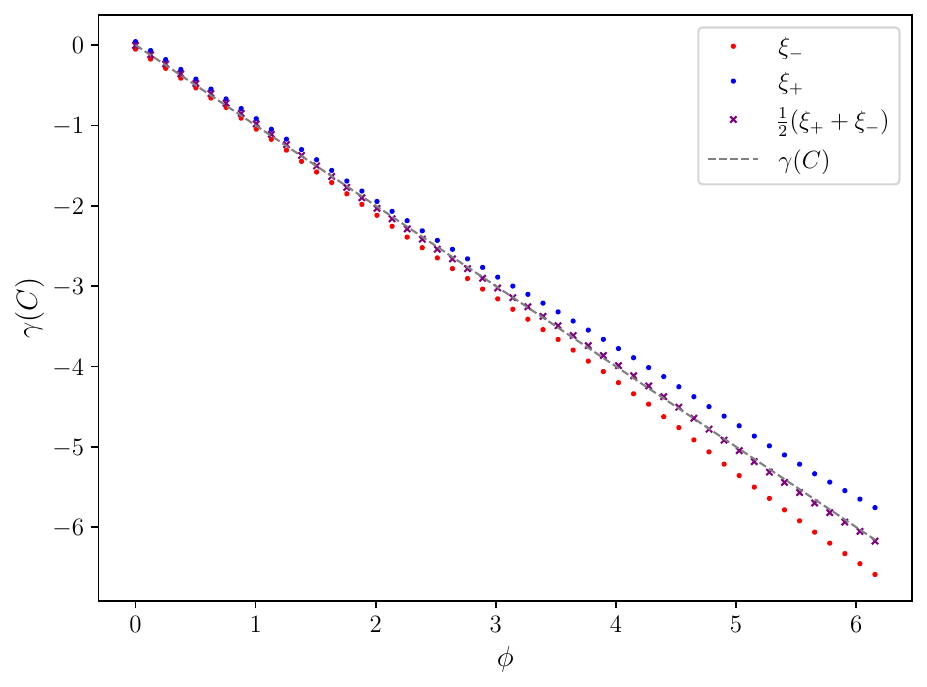}};
          \begin{scope}[x={(img.south east)}, y={(img.north west)}]
              \node at (0.54,-0.05){$\phi_0$};
          \end{scope}
        \end{tikzpicture}
    }\hfill
    \subfloat[\label{fig:Berry's phase graph phi b}]{
        \begin{tikzpicture}
          \node[anchor=south west, inner sep=0] (img)
            {\includegraphics[width=0.48\linewidth,
                trim=0 0.8cm 0 0, clip]{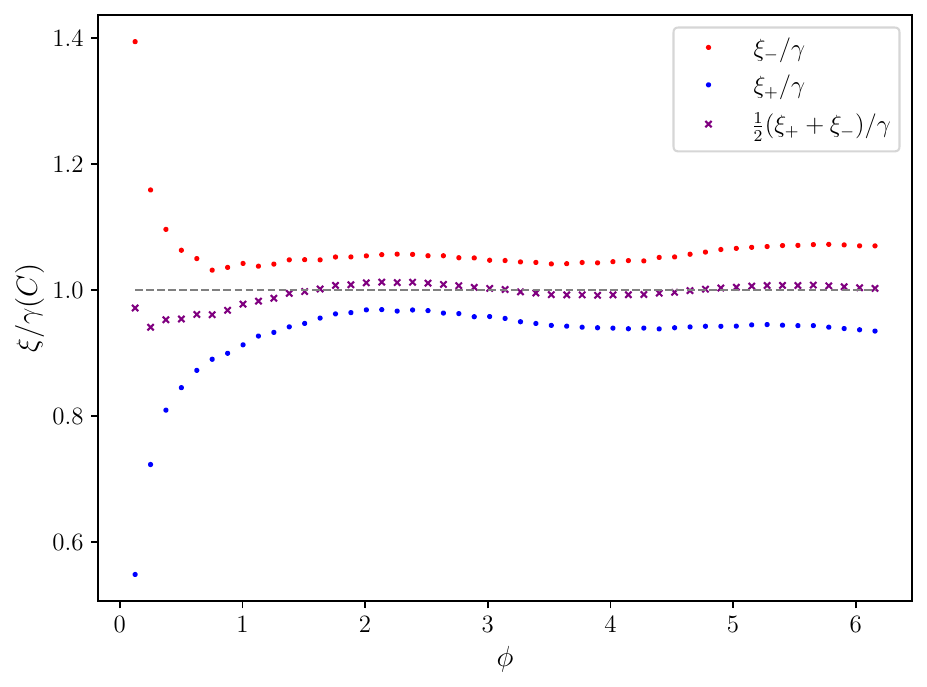}};
          \begin{scope}[x={(img.south east)}, y={(img.north west)}]
              \node at (0.54,-0.05) {$\phi_0$};
          \end{scope}
        \end{tikzpicture}
    }\\[-0.5em]

    \subfloat[\label{fig:Berry's phase graph phi c}]{
        \begin{tikzpicture}
          \node[anchor=south west, inner sep=0] (img)
            {\includegraphics[width=0.48\linewidth,
                trim=0 0.8cm 0 0, clip]{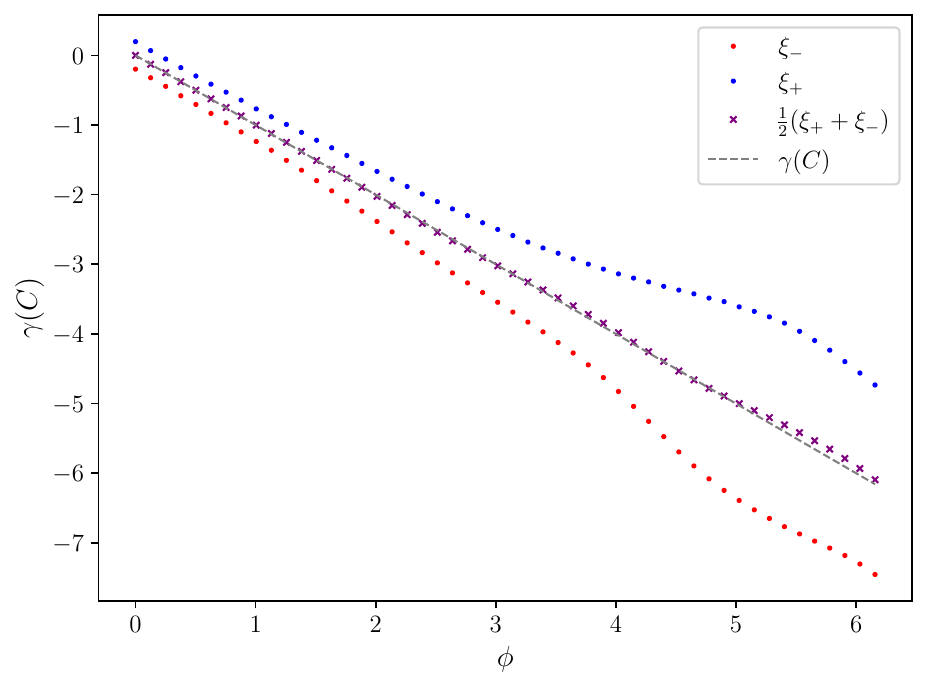}};
          \begin{scope}[x={(img.south east)}, y={(img.north west)}]
              \node at (0.54,-0.05){$\phi_0$};
          \end{scope}
        \end{tikzpicture}
    }\hfill
    \subfloat[\label{fig:Berry's phase graph phi d}]{
        \begin{tikzpicture}
          \node[anchor=south west, inner sep=0] (img)
            {\includegraphics[width=0.48\linewidth,
                trim=0 0.8cm 0 0, clip]{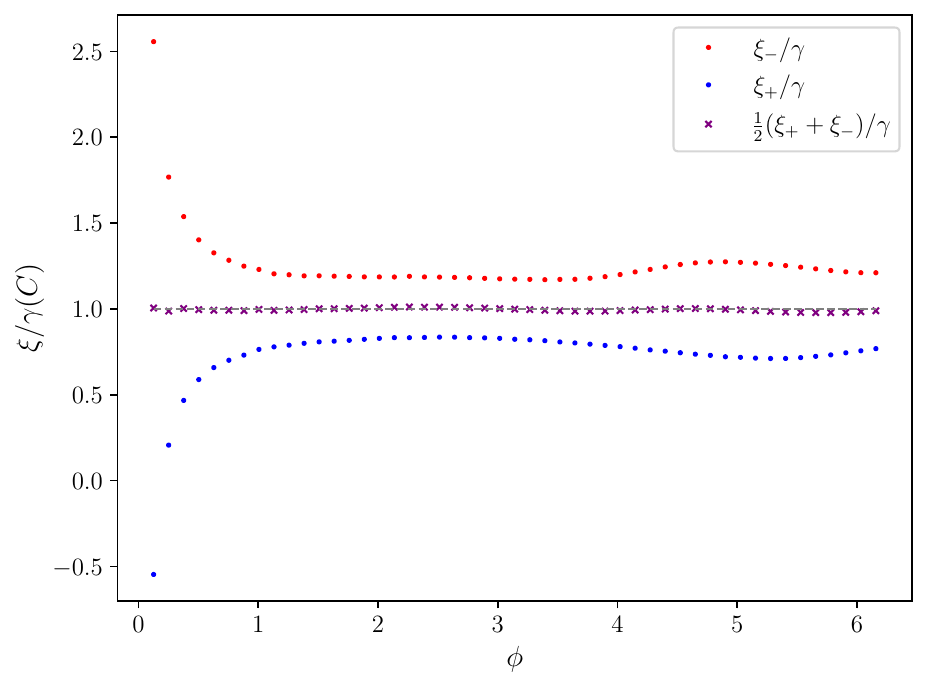}};
          \begin{scope}[x={(img.south east)}, y={(img.north west)}]
              \node at (0.54,-0.05) {$\phi_0$};
          \end{scope}
        \end{tikzpicture}
    }\\[-0.5em]

    \subfloat[\label{fig:Berry's phase graph phi e}]{
        \begin{tikzpicture}
          \node[anchor=south west, inner sep=0] (img)
            {\includegraphics[width=0.48\linewidth,
                trim=0 0.8cm 0 0, clip]{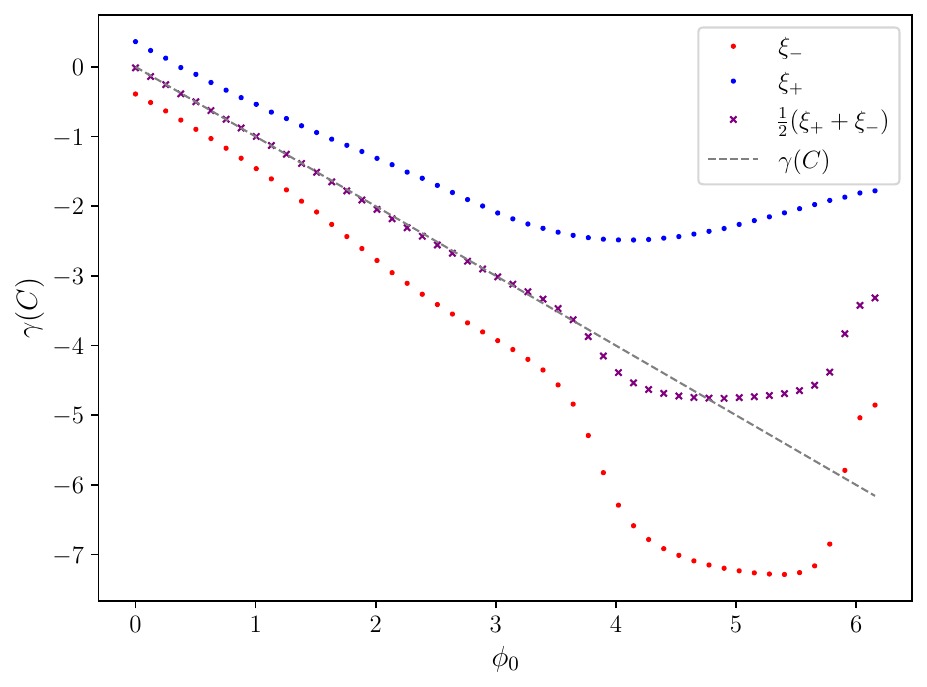}};
          \begin{scope}[x={(img.south east)}, y={(img.north west)}]
              \node at (0.54,-0.05){$\phi_0$};
          \end{scope}
        \end{tikzpicture}
    }\hfill
    \subfloat[\label{fig:Berry's phase graph phi f}]{
        \begin{tikzpicture}
          \node[anchor=south west, inner sep=0] (img)
            {\includegraphics[width=0.48\linewidth,
                trim=0 0.8cm 0 0, clip]{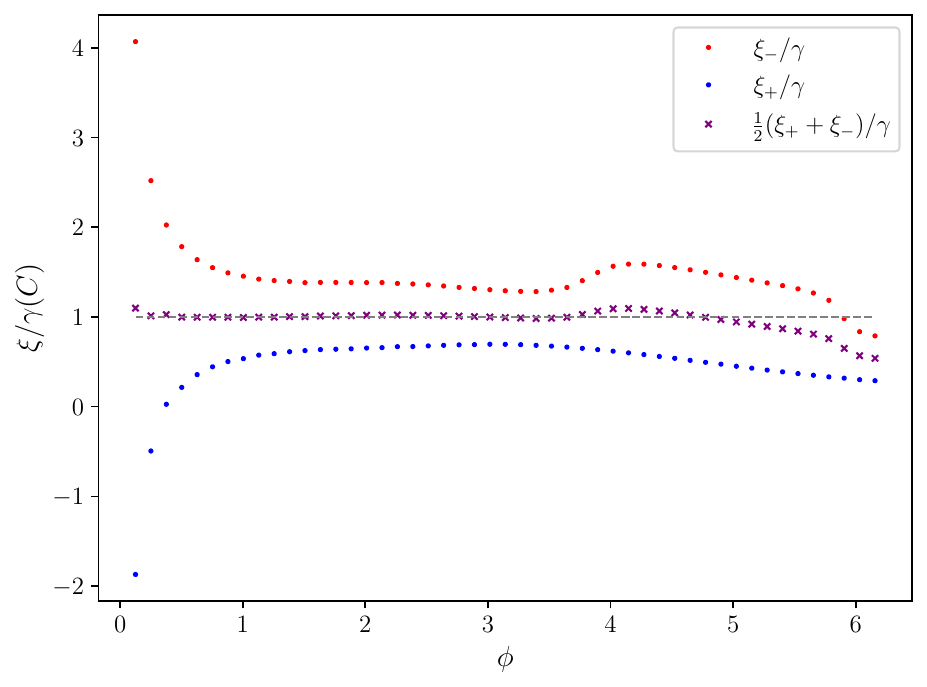}};
          \begin{scope}[x={(img.south east)}, y={(img.north west)}]
              \node at (0.54,-0.05) {$\phi_0$};
          \end{scope}
        \end{tikzpicture}
    }

    \caption{The total phase $\xi_+$ of Eq.~\eqref{eq:AAplus} and $\xi_-$ of Eq.~\eqref{eq:AAminus} for the contour of Fig.~\ref{fig:contour} and its corresponding anti-polar contour respectively. Here we vary $\phi_0$ with $\theta_0=\pi/2$. The ideal Berry's phase $\gamma(C)$ is given in Eq.~\eqref{eq:gamma_C} and indicated by the dashed line. 
    The data points in blue are the $\xi_+$ phases observed when going around the contour in Fig.~\ref{fig:contour}, while those in red are the $\xi_-$ phases for the anti-polar contour.
    In these plots we keep $T_1=T_2=T_3$ for the three segments of the circuit, with total time $T=T_1+T_2+T_3$. In (a) and (b) the total time to complete the circuit is $T=48\pi$. In (c) and (d) the total time $T=12\pi$. Here the error from non-adiabaticity grows with the overall geometric phase, but the averaged Aharonov--Anandan phase still gives the correct $\gamma(C)$. In (e) and (f) the total time is $T=6\pi$ and the averaged Aharonov--Anandan phase fails to give the correct $\gamma(C)$ for $\phi_0 \gtrsim \pi$.}
    \label{fig:Berry's phase graph phi}
\end{figure*}

Finally in this section, let us determine the conditions for adiabaticity in this set-up. 
In particular, in the instantaneous basis of $\vec B\cdot \vec L$ the energy eigenstates are split by $\mu B$ between adjacent $m$ values, and in our initial 
wavefunction in Eq.~\eqref{eq:init} the mode $|0,0,1,0\rangle $ is an $m=0$ momentum eigenstate and has energy $E_{1,1,0}=3$, while the $m=1$ component has energy $3-\mu B$. Therefore our initial state $\ket{\psi(0)}$ is a superposition of two instantaneous energy eigenstates. However, according to the adiabatic theorem, in an adiabatic evolution  mixing between these eigenstates will be exponentially suppressed. The general condition for adiabaticity to hold is 
\begin{align}
    \frac{\braket{m' | \dot H | m }}{(E_{m'}-E_m)^2}~\ll ~1~.
\end{align}
In this particular case we have $E_{m'}-E_m=\mu B$ and $\dot H = {\cal O}(\mu B \alpha /T)$, where $\alpha $ is the total angle moved through by $\vec B$. For the generic single-winding contours considered here,
$\alpha$ is of the same order as the enclosed solid angle and hence the Berry phase $\gamma$. Therefore we expect non-adiabatic couplings to be suppressed provided that  
\begin{align} 
\label{eq:non adiabaticity estimate}
\mu B \, T ~\gg ~\alpha \sim \gamma ~,
\end{align}
where the second approximate equality is a statement about the parameters we shall choose in the study. 

Note that there is a tension between minimising the Trotter error and minimising the non-adiabatic mixing, as the former requires $\mu B dt\ll 1$. If the total number of Trotter steps is $N$, then we require 
\begin{equation}
     N ~\gg ~\frac{1}{\mu B dt} ~\gg ~1~.
\end{equation}

\begin{figure*}[t]
    \centering

    \subfloat[\label{fig:berry_48pi_theta_a}]{
        \begin{tikzpicture}
            \node[anchor=south west, inner sep=0, outer sep=0] (img)
                {\includegraphics[width=0.48\linewidth,
                    trim=0 0.8cm 0 0, clip]{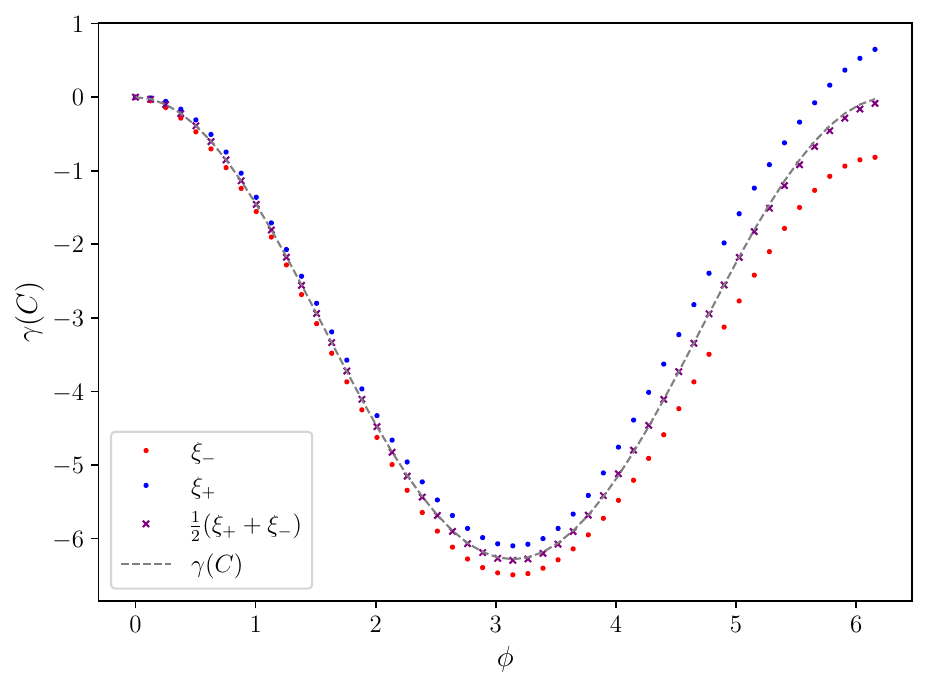}};
            \begin{scope}[x={(img.south east)}, y={(img.north west)}]
                \node at (0.54,-0.05){$\theta_0$};
            \end{scope}
        \end{tikzpicture}
    }\hfill
    \subfloat[\label{fig:berry_48pi_theta_b}]{
        \begin{tikzpicture}
            \node[anchor=south west, inner sep=0, outer sep=0] (img)
                {\includegraphics[width=0.48\linewidth,
                    trim=0 0.8cm 0 0, clip]{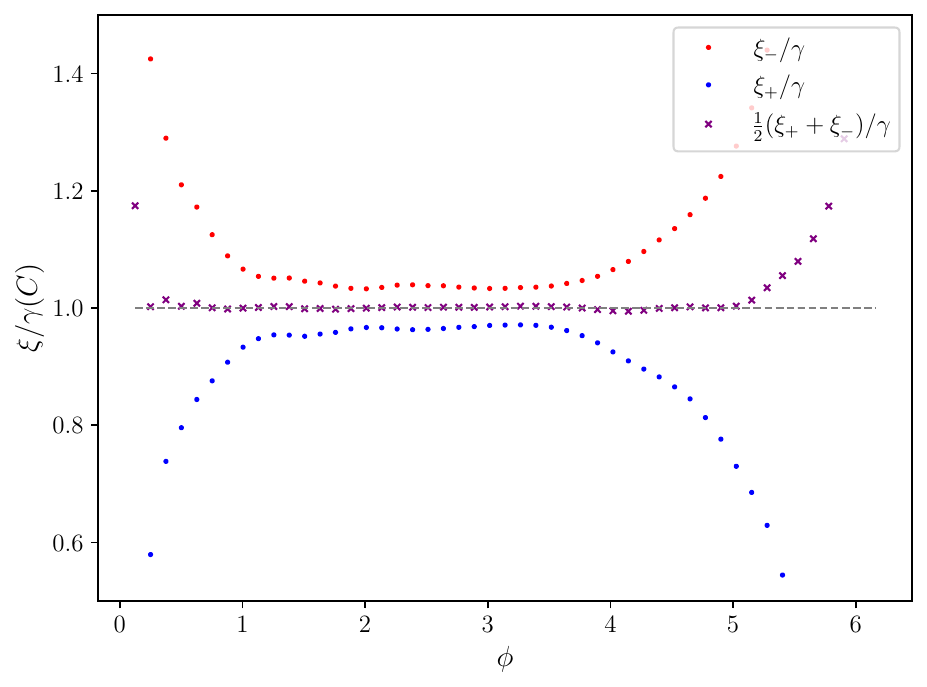}};
            \begin{scope}[x={(img.south east)}, y={(img.north west)}]
                \node at (0.54,-0.05){$\theta_0$};
            \end{scope}
        \end{tikzpicture}
    }

    \caption{As for Fig.~\ref{fig:Berry's phase graph phi a} and \ref{fig:Berry's phase graph phi b} , but with $\phi_0=\pi/2$ and varying $\theta_0$, with the circuit taking total time $T=48\pi$.}
    \label{fig:Berry's phase graph 48pi theta}
\end{figure*}

\section{The Aharonov-Anandan Phase}\label{sec:AAphase}

We now make an important extension of the analysis to include the Aharonov-Anandan phase. This is a generalisation of the Berry phase that is applicable even when the system becomes non-adiabatic. It emerges from any cyclic wavefunction, that is for any state $\ket{\psi}$ that satisfies Schr\"odinger's equation, and is cyclic, \ie, $\ket{\psi(T)}=e^{i\xi}\ket{\psi(0)}$ for some overall phase $\xi\in\Re$ and time period $T\in\Re$. 

In such a situation, the overall phase 
\begin{align}
\xi ~=~ \arg(\braket{\psi(0)|\psi(T)}~,
\end{align}
is still comprised of two distinct contributions, a dynamical phase $\xi_{\rm dyn}$, and a geometric phase, which we will continue to call $\gamma$.
The observation of Aharonov and Anandan was that in quantum mechanics the state vector moves through Hilbert space, but the only physically relevant motion is that of the path traced out by the ray (\ie, the state vector modulo the overall phase), in the {\it projective} Hilbert space. The dynamical part can be {\rm defined} by 
\begin{align}
    \xi_{\rm dyn} ~=~ -\frac{1}{\hbar} \int ^T_0 \braket{H} dt ~=~ - i \int ^T_0 \braket{\psi |\dot\psi } dt~,
\end{align}
using Schr\"odinger's equation. However, this phase is gauge dependent since we may always redefine $\ket{\psi'}=e^{i \chi(t)} \ket{\psi}$ and then $\braket{\psi'|\dot \psi'}=\braket{\psi|\dot \psi}+i\dot\chi$, and we therefore have 
\begin{align} 
i \int ^T_0 \braket{\psi' |\dot\psi' } dt~=~ i \int ^T_0 \braket{\psi |\dot\psi } dt - [\chi(T)-\chi(0)]~.
\end{align}
Aharonov and Anandan note that the difference between the total phase and the dynamical phase, 
\begin{align}
    \gamma ~=~ \xi~+~ i \int ^T_0 \braket{\psi |\dot\psi } dt ~,
\end{align}
clearly {\it is } gauge independent, and has a geometrical interpretation analogous to the Berry phase.  Indeed one can use $\chi$ to entirely remove the dynamical phase for some locally defined 
$\ket{\tilde\psi}$. In this case $\braket{\tilde \psi |\dot{ \tilde \psi} }=0$ trivially, and the entire phase $\xi $ becomes equal to just the geometric phase, $\gamma$. As with the Berry phase, this remaining geometric phase is the only physically meaningful one. 

The system we are studying is of particular interest because it allows us to isolate the phase $\gamma$ as follows. Suppose that we begin in state $\phi_{0,1,1} $ and take the system non-adiabatically around the trajectory $\vec{B}(t)$ defined in Eq.~\eqref{eq:angles_T}. The total phase $\xi_+$ is measured as the phase $\Delta$ in Eq.~\eqref{eq:final}. That is, we would measure 
\begin{align}
    \Delta ~=~ \xi_+ ~=~ \mu B f(T) + \gamma(C) ~.
    \label{eq:AAplus}
\end{align}
Here $f(T)$ is some function that encodes non-adiabatic effects such as Landau-Zener mixing, where the original state leaks to other eigenstates, and also leading Trotter error (which can itself be thought of as a form of non-adiabaticity). Thus $f(T)$ tends to $T$ when the system becomes adiabatic: that is  $\lim_{T\to\infty } f(T)/T = 1$.

We then repeat the experiment with a reversed magnetic field, $\vec{B}(t)\to -\vec{B}(t)$, and find a phase $\xi_-$. In the second, mirror, case the state $\phi_{0,1,1} $ is an excited state. However, the splitting of energy levels does not change (and in particular the energy levels do not cross). Therefore we do not expect to generate Landau-Zener transition, but rather this evolution will also have a Landau-Zener mixing function. It turns out that to first order (for reasons we explain in a moment), even when $\Delta$ is large this function is the same as $f(T)$, such that 
\begin{align}
   \Delta ~=~ \xi_- ~=~ - \mu B f(T) + \gamma(C) ~.
   \label{eq:AAminus}
\end{align}
The geometric phase can then be determined as 
\begin{align} 
\label{eq:avgeG}
\gamma(C) ~=~ \frac{1}{2} (\xi_++\xi_-)~,
\end{align} 
where $\gamma(C)$ is the Berry phase in Eq.~\eqref{eq:gamma_C}.

Why would we expect to find the same $f(T)$ function with the reversed magnetic field, even when the overall phase $\Delta$ is large? The reason for this is the particular three-spin configuration that we are considering. To make this explicit note that the Hamiltonian may be written as 
\begin{align}
\label{eq:Hpm}
H~\supset ~ -\mu \vec L \cdot \vec B ~=~ \mu \left(  L_z B_z + L_+ B_- + L_- B_+\right)
\end{align}
with $L_\pm $ being the usual ladder operators for azimuthal angular momentum. 
When we perform the experiment with reflected $\vec B$, according to Eq.~\eqref{eq:E01m} it is exactly equivalent to considering the $m=-1$ state with the $\vec B$ field un-reflected. For $\ell =1$ we have $L_\pm\ket{m=\mp} = \ket{m=0} $ and hence crucially $\braket{m=1|L_\pm |m=-1}=0$. This implies that the leading mixing effect is from the $m=1$ state to the  $m=0$ state in the original system, and from the $m=-1$ state to the  $m=0$  state in the reflected system. Meanwhile the $m=1$ and $m=-1$ states do not directly mix into each other. Thus, to leading order there is no asymmetry in the transition probabilities to the neighbouring $m=0$ state in the original and reflected cases.

This provides us with several checks. We will of course be able to determine the Berry phase even without adiabaticity, and we expect to reproduce results approaching Eq.~\eqref{eq:gamma_C}. But thanks to the separate $\xi_+$ and $\xi_-$ branches, it will also allow the observation of Landau-Zener mixing and the asymptote to adiabaticity as we increase $T$. 

In addition it allows us to test where the first-order Landau-Zener mixing assumption breaks down, giving a minimal circuit for consistent measurements of the phenomenon. Where do we expect this first-order assumption, and hence Eq.~\eqref{eq:avgeG}, to fail? The leading contribution to $f(T)$ comes from terms linear in $H=-\mu\vec L \cdot \vec B$ giving contributions  $\Delta\approx -\mu B f(T) $. The first subleading term is from the double-hop from the excited $m=-1$ state to the $m=+1$ state. This amplitude is of order $ \frac{\alpha^2}{(\mu B T)^2}$. This is to be compared with the leading single-hop term $\alpha/(\mu B T) $ which implies a breakdown of our averaged Aharonov-Anandan approximation precisely where Eq.~\eqref{eq:non adiabaticity estimate} predicts loss of adiabaticity. In other words the parameter \begin{align}
\label{eq:varepsi}
    \varepsilon ~=~ \frac{\alpha}{\mu B T}
\end{align} governs the accuracy of our predictions, but the $\gamma $ that we deduce from the average in Eq.~\eqref{eq:avgeG} is in fact accurate to order ${\cal O}(\varepsilon^2)$: in summary then, we expect to see $\gamma(C) ~=~ \xi_+-{\cal O}(\varepsilon)
$ and $\gamma(C) ~=~ \xi_- + {\cal O}(\varepsilon)
$, but 
\begin{align} 
\label{eq:avgeG2}
\gamma(C) ~=~ \frac{1}{2} (\xi_++\xi_-)~+~{\cal O}(\varepsilon^2)~.
\end{align} 

\begin{figure*}[t]
    \centering
    \subfloat[\label{fig:qpu_results_a}]{
        \centering
        \begin{tikzpicture}
          \node[anchor=south west, inner sep=0] (img)
            {\includegraphics[width=0.48\linewidth,
                trim=0 0.8cm 0 0, clip]{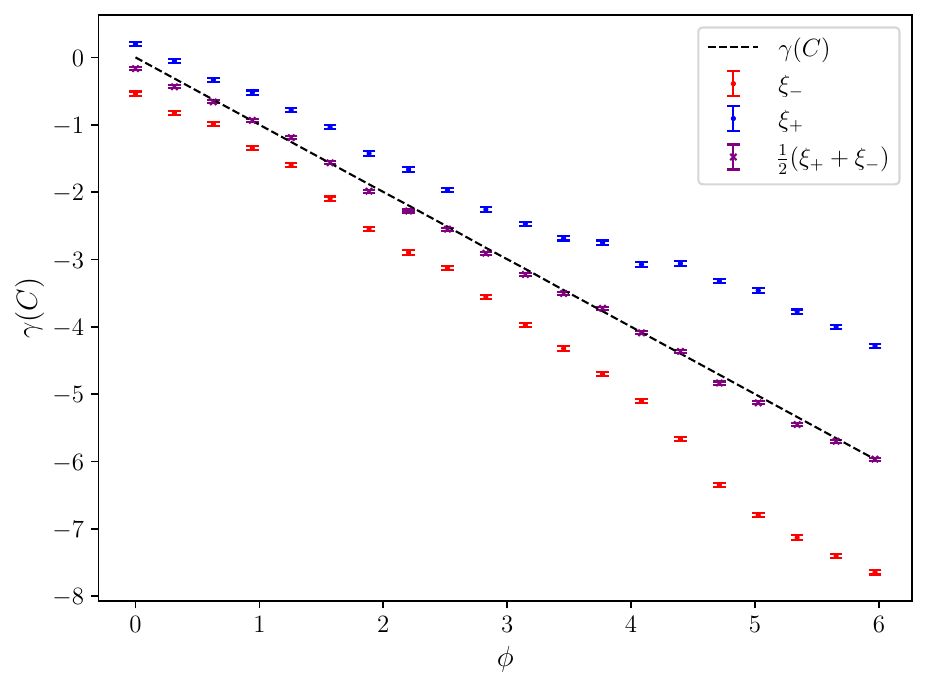}};
          \begin{scope}[x={(img.south east)}, y={(img.north west)}]
              \node at (0.54,-0.05){$\phi_0$};
          \end{scope}
        \end{tikzpicture}
    }\hfill
    \subfloat[\label{fig:qpu_results_b}]{
        \centering
        \begin{tikzpicture}
          \node[anchor=south west, inner sep=0] (img)
            {\includegraphics[width=0.48\linewidth,
                trim=0 0.8cm 0 0, clip]{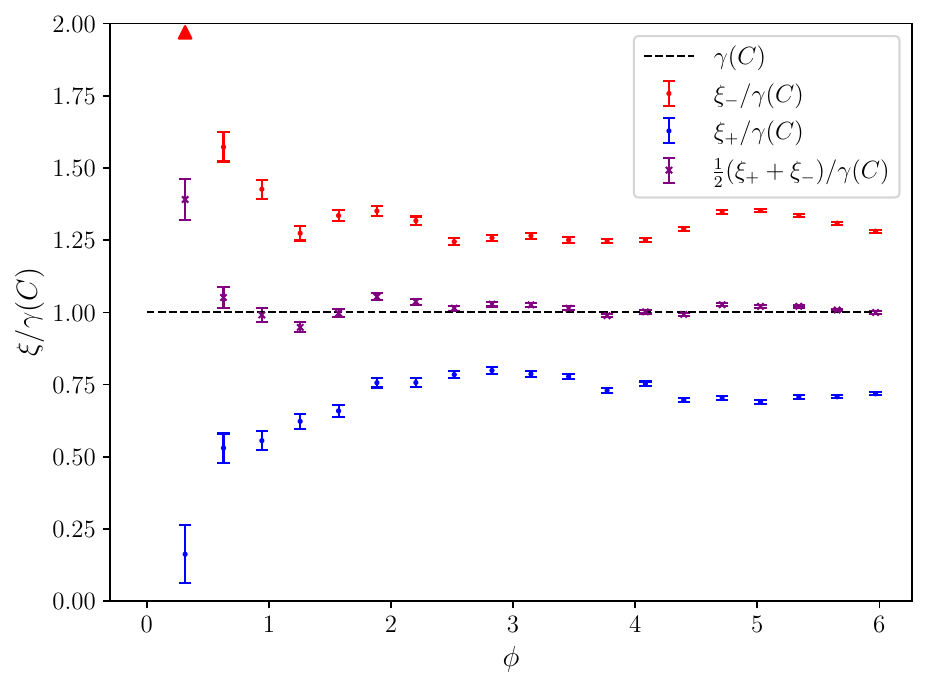}};
          \begin{scope}[x={(img.south east)}, y={(img.north west)}]
              \node at (0.56,-0.05) {$\phi_0$};
          \end{scope}
        \end{tikzpicture}
    }
    \caption{Results from the {\tt Quandela Ascella} quantum computer. The total phase $\xi_+$ of Eq.~\eqref{eq:AAplus} and $\xi_-$ of Eq.~\eqref{eq:AAminus} for the contour of Fig.~\ref{fig:contour} and its corresponding anti-polar contour respectively. Here we vary $\phi_0$ with $\theta_0=\pi/2$, and take total time $T=8\pi$ to complete the circuit with $T_1=T_3=2\pi$ and $T_2=4\pi$. The ideal Berry's phase $\gamma(C)$ is given in Eq.~\eqref{eq:gamma_C} and indicated by the dashed line. The data points in blue are the $\xi_+$ phases observed when going around the contour in Fig.~\ref{fig:contour}, while those in red are the $\xi_-$ phases for the anti-polar contour.}
    \label{fig:qpu_results}
\end{figure*}

\section{Results}\label{sec:results}

\subsection{Simulation on {\tt Quandela Perceval}}

In order to test our analysis above, the circuit was first implemented on the {\tt Quandela Perceval} simulator \cite{heuser2022perceval}, which is an open-source Python framework. {\tt Perceval} provides state-vector and sampling-based backends, with Symbolic Linear Optical Simulator (SLOS) exact solvers.

Using the simulator the circuit was run for the contour of Eq.~\eqref{eq:gamma_C},
with $\mu B=1$ over various times $T=T_1+T_2+T_3$,  first varying $\phi_0$ with fixed $\theta_0$, and then vice-versa. In these runs we take $T_1=T_2=T_3$ for the three segments of the contour. 

Fig.~\ref{fig:Berry's phase graph phi} shows the measured Aharonov-Anandan phases (\ie~ the values of $\xi_\pm$ and their average) for fixed 
$\theta_0=\frac{\pi}{2}$ and varying $\phi_0$, which has Berry phase $\gamma(C)=-\phi_0$ for this circuit. The theoretical adiabatic ideal $\gamma$ is shown as a dashed line. Note that in all our Figures we will take a fixed Trotter step of $dt = 0.01\times \pi$. 

Figs.~\ref{fig:Berry's phase graph phi a} and \ref{fig:Berry's phase graph phi b} show the results for $T=48\pi$.
Clearly both $\xi_+$ and $\xi_-$ are very close to the theoretical Berry phase, but what is notable is that, while they individually diverge from the theoretical Berry phase as the cycle grows, their average follows its theoretical value almost exactly, regardless of $\phi_0$. This is as predicted by Eq.~\eqref{eq:avgeG2}, since for this choice of parameters we would estimate $\varepsilon \approx \phi_0/T_2 = \phi_0/16\pi$, taking the $T_2$ segment as representative of the whole evolution. When $\phi_0=2\pi$ we expect an error of order ${\cal O}(1/64)\sim 1\%$ in the $\gamma $ that we deduce from the averaged Aharonov-Anandan phase. 

In Figs.~\ref{fig:Berry's phase graph phi c} and \ref{fig:Berry's phase graph phi d} we show an example where the circuit is completed in a much shorter overall time ($T=12\pi $). We see that the behaviour is now  far from adiabatic for large $\phi_0$ since we estimate $\varepsilon \sim 1/2$ at the largest values of $\gamma$. Here we see a roughly 25\% deviation in the $\xi_\pm$ values from $\gamma$ in Fig.~\ref{fig:Berry's phase graph phi d}, but with the average in  Eq.~\eqref{eq:avgeG2}
still reproducing $\gamma(C)$ to an accuracy of order 5\%, showing its intrinsic robustness. 
As the Trotter error must be the same as in Fig.~\ref{fig:Berry's phase graph phi a} and \ref{fig:Berry's phase graph phi b} we see that the non-adiabaticity is providing the dominant error. 

In Figs.~\ref{fig:Berry's phase graph phi e} and \ref{fig:Berry's phase graph phi f} we show a run in which the circuit was completed in total time $T=6\pi$. Here we see that the $\xi_-$ curve fails to mirror the $\xi_+$ curve for values of $\phi_0\gtrsim \pi$, and the  averaged Aharonov-Anandan phase no longer adheres to the geometric phase $\gamma(C)$. Here we have $\varepsilon = \phi_0/2\pi$ and thus adiabaticity is indeed expected to be lost completely at larger values of $\phi_0$, with Eq.~\eqref{eq:avgeG2} failing at this point.

Finally in Fig.~\ref{fig:Berry's phase graph 48pi theta} we present the results for the same circuit with total time $T=48\pi$ but varying $\theta_0$ with $\phi_0=\pi/2$, for which we see the expected form of geometric phase, namely $\gamma = \cos\theta_0 - 1$.

As expected we do not see any evidence of Landau-Zener transition in any of these graphs.

\subsection{Implementation on {\tt Quandela Ascella} QPU}

To further validate our model, we exploit the exact correspondence between passive Gaussian operations in the continuous-variable framework and the linear-optical unitaries implemented by single-photon photonic processors. This allows for the experimental realisation of the circuit on the {\tt Quandella Ascella} platform, a photonic quantum computer operating a {\tt MOSAIQ-6} processor: a reconfigurable 12-mode interferometer capable of supporting up to six simultaneously injected single photons. Since our algorithm requires only beam splitters and phase rotations, the entire Trotterised evolution of the three-segment contour of Fig.~\ref{fig:contour} can be executed natively on the device. 

The system was prepared in the initial state outlined in Section~\ref{sec:state_prep} by injecting a single photon into the $x$-mode. Time evolution has been performed for a total time of $T=8\pi$ and a Trotter time step size of $dt = 0.1\pi$, with the evolution along the three contour segments partitioned as $(T_1, T_2, T_3)=(2\pi, 4\pi, 2\pi)$. We vary $\phi_0$ while keeping $\theta_0 = \pi/2$. The Aharonov-Anandan phases, $\xi_\pm$, have been extracted from the device with each data point obtained from 1000 shots; statistical uncertaintites have been included~\footnote{Note, there is some ambiguity in the transpilation for $\phi_0 = \pi$ exactly, thus a value of $\phi_0 = \pi + 0.005$ has been used.}. The results from the quantum computer are shown in Fig.~\ref{fig:qpu_results}.

Figure~\ref{fig:qpu_results_a} shows the raw geometric phases, $\xi_\pm$, obtained from {\tt Ascella} compared to the theoretical adiabatic ideal value for $\gamma$, indicated by the black dashed line. For small and intermediate values of $\phi_0$, both $\xi_+$ and $\xi_-$ remain close to the theoretical value of the Berry phase, $\gamma(C) = -\phi_0$, indicating that the evolution remains in the adiabatic regime despite the small evolution time of $T= 6\pi$. As $\phi_0$ increases, the two phase curves diverge from the adiabatic prediction, following the same trend observed in Figs.~\ref{fig:Berry's phase graph phi c} and \ref{fig:Berry's phase graph phi d}.

Despite the deviations of the individual phases, the averaged Aharonov–Anandan phase, $\frac{1}{2}\left(\xi_+ + \xi_-\right)$, remains in good agreement with $\gamma(C)$, demonstrating that the quantum device is capable of experimentally capturing the cancellation of the dynamics phases. The normalised data shown in Fig.~\ref{fig:qpu_results_b} reinforce this conclusion. While $\xi_\pm/\gamma(C)$ show significant deviations at large $\phi_0$, the averaged quantity remains comparatively stable and close to unity until the onset of the strongly non-adiabatic regime. 

Overall, the {\tt Ascella} device performs remarkably, reproducing the key qualitative features observed in the quantum emulation, including the robustness of the averaged Aharonov-Anandan phase, and the characteristic divergence of the individual phases at large values of $\phi_0$. These results confirm that our model can be faithfully implemented on current photonic quantum hardware using only passive linear optics, and that the cancellation of non-geometric phases is experimentally accessible even in regimes where non-adiabatic effects are substantial.

\section{Conclusions}\label{sec:conclusions}

In this paper, we have presented a continuous-variable quantum computing (CVQC) framework for studying Berry’s phase on photonic quantum architectures, encoding information in the quadrature operators of the simple harmonic oscillator, $\hat x$ and $\hat p$. We demonstrated that the CVQC framework naturally captures the time evolution of a charged particle with orbital angular momentum subjected to an adiabatically varying magnetic field. Moreover, we showed that this evolution can be realised using entirely passive linear optics: a regime in which continuous-variable Gaussian operations coincide exactly with the linear-optical unitaries of single-photon platforms. Although our construction is developed within the CVQC formalism, this equivalence enables a direct, one-to-one mapping onto existing single-photon photonic devices, permitting experimental realisation on the {\tt Quandela Ascella} quantum processor.

Using this formulation, we simulated the real-time evolution on a quantum emulator and examined how the geometric phase behaves across different dynamical regimes. The simulated results show excellent agreement with the theoretical Berry phase in the adiabatic limit. Furthermore, the averaged Aharonov-Anandan phase was found to robustly suppress non-adiabatic dynamical contributions, continuing to track the geometric phase even when far into the non-adiabatic regime. The model was implemented on the {\tt Quandela Ascella} device, a single-photon quantum computer that realises arbitrary passive linear-optical transformations. The hardware results closely match the emulator, demonstrating that the extraction of the geometric phase is experimentally possible even when non-adiabatic effects are substantial on current photonic quantum devices. 

Overall, this work establishes a practical bridge between CVQC formulations and present-day linear-optical hardware. It highlights that geometric and topological phase phenomena, often formulated in continuous-variable settings, can be explored experimentally on existing single-photon architectures without the need for non-Gaussian gate operations. The methods developed here open the door to extending geometric-phase techniques to higher-dimensional orbital-angular-momentum spaces, non-adiabatic geometric gates, and potentially holonomic protocols on photonic quantum computers.

\begin{acknowledgments}
This work was supported by the STFC under IPPP grant no. ST/T001011/1, and by the London Mathematical Society under grant no. URB-2025-25, and the Department of Mathematical Sciences in Durham. All scientific content, analyses, and conclusions were generated and verified by the authors, who take full responsibility for the manuscript.
\end{acknowledgments}

\bibliographystyle{inspire}
\bibliography{refs}{}

\end{document}